\documentclass[10pt]{article}

\usepackage{amsmath}
\usepackage{amssymb}

\usepackage{graphicx}

\usepackage{cite}

\usepackage{color} 

\usepackage[labelfont=bf,labelsep=period,justification=raggedright]{caption}

\makeatletter
\renewcommand{\@biblabel}[1]{\quad#1.}
\makeatother

\date{}

\newcommand{\er}{Erd\"os-R\'enyi}
\newcommand{\be}{\begin{eqnarray}}
\newcommand{\ee}{\end{eqnarray}}
\newcommand{\la}{\langle}
\newcommand{\ra}{\rangle}
\begin{document}

\begin{flushleft}
{\Large 
\textbf{Modularity and Anti-Modularity in Networks With Arbitrary Degree Distribution}
}
\\
Arend Hintze$^{1}$, 
Christoph Adami$^{2,\ast}$ 
\\
\bf{1} Keck Graduate Institute of Applied Life Sciences, 535 Watson Drive, Claremont, CA 91711
\\
$\ast$ E-mail: adami@kgi.edu
\end{flushleft}

\section*{Abstract}
Much work in systems biology, but also in the analysis of social network and communication and transport infrastructure involves an in-depth analysis of local and global properties of those networks, and how these properties relate to the function of the network within the integrated system. Most often, systematic controls for such networks are difficult to obtain, because the features of the network under study are thought to be germane to that function. 
In most such cases, a surrogate network that carries any or all of the features under consideration, but is created artificially and therefore carries none of the function of the network being studied, would be of considerable interest. Here, we present an algorithmic model for growing networks with arbitrary degree distributions and arbitrary modularity using a small set of parameters. We show that the degree distribution is controlled mainly by the ratio of node to edge addition probabilities, and the probability for node duplication. We compare topological and functional modularity measures, study their dependence on the number and strength of modules, and introduce the concept of anti-modularity: a property of networks in which nodes from one functional group preferentially do not attach to other nodes of that group. We also investigate global properties of networks as a function of the network's growth parameters, such as smallest path length, correlation coefficient, small-world-ness, and the nature of the percolation phase transition. We search the space of networks for those that are most like some well-known biological examples, and analyze the biological significance of the parameters that gave rise to them. We find that many of the celebrated network properties may be a consequence of the way in which they grew, rather than a necessary consequence of how they work or function.


\section*{Author Summary}
Networks describing the interaction of the elements that constitute a complex system grow and develop via a number of different mechanisms, such as the addition and deletion of nodes, the addition and deletion of edges, as well as the duplication or fusion of nodes. 
While each of these mechanisms can have a different cause depending on whether the network is biological, technological, or social, their impact on the network's structure, as well as its local and global properties, is similar. This allows us to study how each of these mechanisms affects networks either alone or together with the other processes, and how they shape the characteristics that have been observed.  We study how a network's growth parameters impact the distribution of edges in the network, how they affect a network's modularity, and point out that some parameters will give rise to networks that have the opposite tendency, namely to display {\em anti-modularity}. Within the model we are describing, we can search the space of possible networks for parameter sets that generate networks that are very similar to well-known and well-studied examples, such as the brain of a worm, and the network of interactions of the proteins in baker's yeast.

\section*{Introduction}

The representation of complex interacting systems as networks has become commonplace in modern science~\cite{Barabasi2002,BornholdtSchuster2002,Watts2003,Newmanetal2006,Barratetal2008}. While such a representation in terms of nodes and edges is near-universal, the systems so described are highly diverse. They range from biological (e.g., protein interaction graphs, metabolic reaction networks, neuronal connection maps) over engineering (blueprints, circuit diagrams, communication networks) to social systems (friends, collaboration, or citation networks). One of the hallmarks of human-designed systems appears to be their modularity~\cite{ClarkBaldwin2000}: systems designed in a modular fashion are more robust to component failure, can be quickly repaired by switching out defective modules, and their designs are easier to understand for a human engineer. Systems that emerged via biological evolution rather than design do not have to be easily understandable, but robustness and repair are still important characteristics. Beyond those, it appears that biological systems need to be {\em evolvable}~\cite{KirschnerGerhart1998,Wagner2005a,Wagner2005b}. While this criterion appears circular because obviously biological systems {\em have} evolved, there are differences in the degree of evolvability, which determine how well a system can adapt to changing environments. Modularity has been identified as possibly a key ingredient in evolvability, because it can both supply mutational robustness via the isolation of components and fast adaptation via the recombination of parts, or by altering the connections between the modules~\cite{Hartwelletal1999,SchlosserWagner2004,CallebautRasskin2005,Wagner2005a,Alon2007}. While our intuitive understanding of modularity is simple (from a designer's point of view) as ``discrete entities whose function is separable from those of other modules"~\cite{Hartwelletal1999}, the identification of modules from a representation of the system as a network is not straightforward. Commonly, modules in networks are identified via clustering algorithms that identify groups of strongly interconnected nodes that are only weakly connected to other such nodes~\cite{RivesGalitski2003,SpirinMirny2003,Clausetetal2004,Newman2006}, but often information external to the purely topological structure is used to determine modular relationships, such as co-regulation~\cite{Segaletal2003,Segaletal2004} or evolutionary conservation~\cite{Sneletal2002,Qinetal2003,Slonimetal2006}. 
When the modular or community structure of a network is given or known, different measures exist to quantify the {\em extent} of modularity in the network~\cite{Guptaetal1989,Newman2002,Newman2003b,NewmanGirvan2004,Middendorfetal2005,Newman2006}.

Another defining characteristic of networks is their edge (or degree) distribution: the probability $p(k)$ that a randomly picked node of the network has $k$ edges. Regular graphs, for example, are networks where each node has exactly the same number of edges as any other (a square lattice is a regular graph of degree four, except for the edge and corner nodes). Graphs can also be constructed randomly, by adding edges between nodes with a fixed probability. The first description of the connectivity distribution of such random graphs is due to Erd\"os and R\'enyi~\cite{ErdosRenyi1959,ErdosRenyi1960,ErdosRenyi1961} and Solomonoff and Rapoport~\cite{SolomonoffRapoport1951}. These authors found that the distribution of edges in such graphs is binomial, or,  in the limit of a large number of nodes, approximately Poisson. While random networks can be found in social interaction and engineering networks~\cite{Amaraletal2000}, they are comparatively rare in nature. For example, 
the edge distribution of the only biological neural network mapped to date (the brain of the nematode {\it C. elegans})~\cite{Whiteetal1986} is consistent with that of an \er\ network~\cite{Amaraletal2000,Reigletal2004}. 

Most other networks found in nature, however, have a {\em scale-free} edge distribution, implying that just a few nodes have very many edges, while most nodes are connected to only a few. The emergence of this scale-free degree distribution can be understood in many different ways~\cite{BarabasiAlbert1999,AlbertBarabasi2000,Vazquez2003,BaukeSherrington2007} (see also~\cite{Pfeifferetal2005,BornholdtRohlf2000}) and usually requires a {\em growth process} where either nodes with many edges preferentially attach to other nodes with many edges, or else grow via node duplication and mutation~\cite{Raval2003}  (see~\cite{Newman2003} for a review of growth models). Indeed, graphs obtained by a growth process appear to show preferential attachment naturally~\cite{Vazquez2003} (because the oldest nodes usually have more edges than younger nodes) and  are fundamentally different from those produced probabilistically~\cite{Callawayetal2001}.

Here, we study the processes giving rise to particular edge distributions in detail, using a simple algorithm that can produce networks with any degree distribution and any modularity. In particular, we generate graphs with defined functional modules using an assortativity matrix, and study how modularity depends on a number of different parameters. We also  introduce a new measure of functional modularity that only takes into account whether or not nodes that have been assigned to the same functional group connect to each other. Using this measure, we can show that some classes of networks can be {\em anti-modular}, that is, they show a tendency of nodes with the same functional assignment {\em not} to be connected to each other. Finally, we use the network growth model to investigate global properties of networks, and study the set of parameters giving rise to networks similar to well-known biological networks.

\section*{Results and Discussion}

A fundamental difference between random graphs and networks with scale-free edge distribution is thought to be the way in which these networks are generated. Random networks are created by randomly adding edges between existing nodes, while scale-free networks are usually generated by growth via preferential attachment~\cite{BarabasiAlbert1999,Price1976,Dorogovtsevetal2000,Callawayetal2001,Krapivskyetal2000} or else grown via duplication with subsequent diversification~\cite{Raval2003,PastorSatorras2003,Bebek2006}. For some networks (mostly metabolic reaction networks~\cite{FellWagner2000,Jeongetal2003})  preferential attachment is not sufficient to explain their degree distribution~\cite{Newman2003a}. Here we describe an algorithm that will grow networks with any degree distribution based on a growth model with only a few parameters, with or without duplication. Depending on those parameters, we can obtain Erd\"os-R\'enyi-like graphs, networks with scale-free degree distribution, small-world networks, regular graphs and lattices, bi-partite and $k$-partite graphs, and anything in between. In addition, this algorithm is able to grow those networks with any degree of modularity and arbitrary size. The growth parameters can even be chosen in such a way that the resulting networks actually show a negative modularity score, that is, they can be {\em anti-modular}. Here, we give a very brief overview of the model's parameters, which are described in more detail in {\bf Models}. 

The main parameters of the growth model are summarized in Table 1. At each {\em event}, a node is added with probability $pP_N$ (while a node is removed with probability $(1-p)P_N$), and an edge is added (removed) with probability $qP_E$ [$(1-q)P_E$]. At the same time, a node is duplicated (fused) with probability $rP_D$ [$(1-r)P_D$]. Modules are built from an assortativity matrix (for $N_c$ modules $M_1,...,M_{N_c}$)
\be \label{ematrix}
 e=\left(  \begin{matrix} 
      p(M_1\rightarrow M_1) & p(M_1\rightarrow M_2)& \ldots& p(M_1\rightarrow M_{N_c}) \\
     p( M_2\rightarrow M_1) & p(M_2\rightarrow M_2)& \ldots& p(M_1\rightarrow M_{N_c}) \\
        \vdots&\vdots&\ddots& \vdots\\
      p(M_{N_c}\rightarrow M_1)&  p(  M_{N_c}\rightarrow M_2)& \ldots& p(  M_{N_c}\rightarrow M_{N_c})\\
   \end{matrix}\right)\;.
\ee

\subsection*{Growing scale-free and other networks}
The standard model for growing random graphs is due to Callaway et al.~\cite{Callawayetal2001}, who introduced a model where a node is added at each event, and an edge is added with a given probability per event. While there are no duplications in this model and edges are not preferentially attached to high-degree nodes, there is still a form of preferential attachment because older nodes have a higher probability of obtaining edges, and also have a higher probability of already sporting more edges~\cite{Callawayetal2001}. This model produces exponential degree distributions whose form can be predicted exactly, but scale-free distributions cannot be produced. With the present network growth model, it is easy to grow networks with arbitrary degree distribution, by changing just a few parameters. 

\begin{figure}[htbp] 
   \centering
   \includegraphics[width=3.5in]{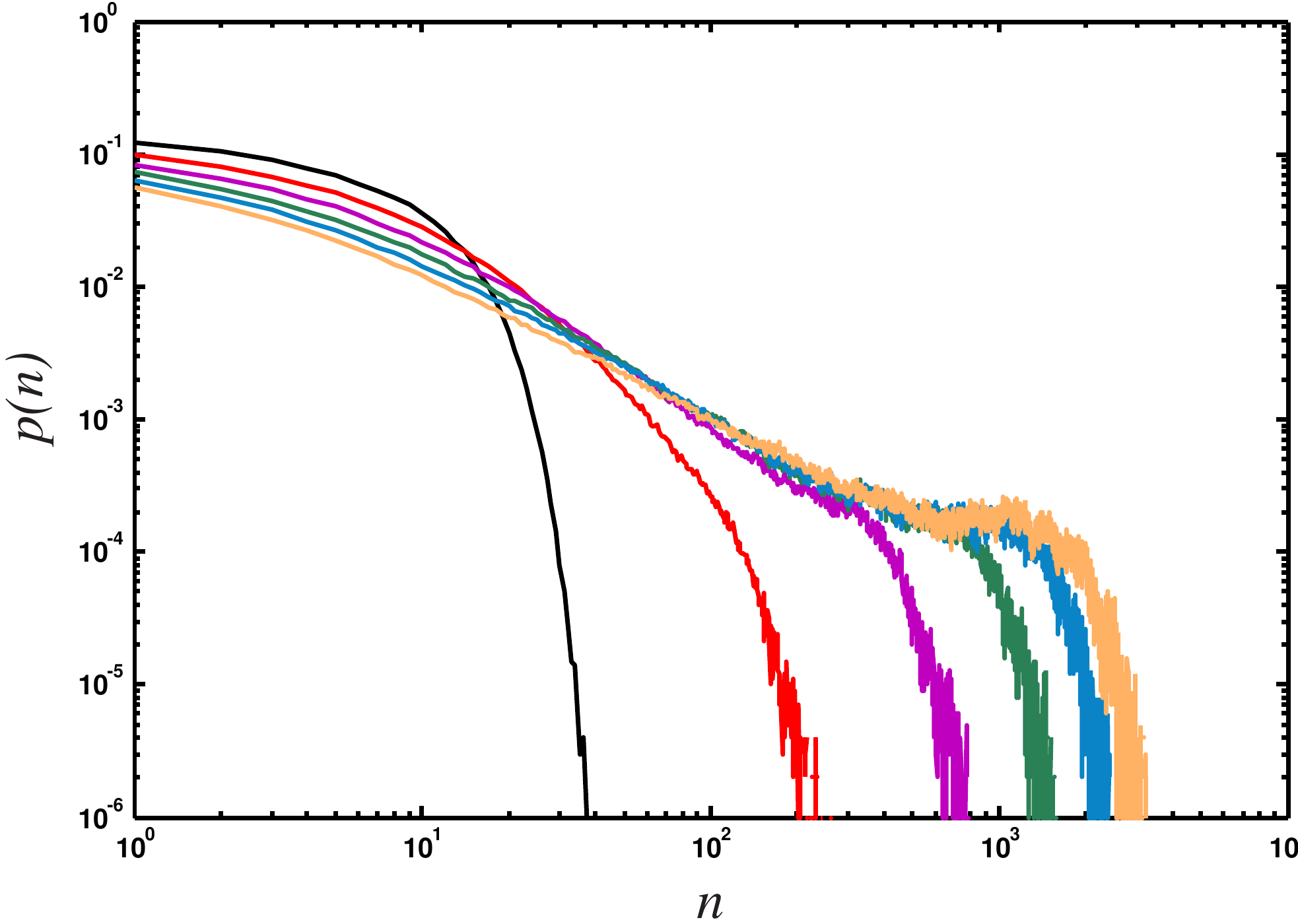} 
   \caption{{\bf Degree distribution as a function of duplications.} The degree distribution of randomly grown networks with different node duplication probabilities $P_D$ ($r=1$), at fixed $P_N=0.2$, $P_E=0.75$ (with $p=1,q=1$). $P_D=0$ (black), $P_D=0.1$ (red), $P_D=0.2$ (magenta), $P_D=0.3$ (green), $P_D=0.4$ (blue), and $P_D=0.5$ (yellow). Average of 100 replicates of networks grown to size $n=$1,000.}
   \label{fig:degdist}
\end{figure}

To begin with, we test whether the exponential distribution of the Callaway model morphes into a scale-free distribution as the duplication event probability is increased. In Fig.~\ref{fig:degdist} we show the degree distribution with a fixed node and edge event probabilities, but changing the node duplication probability $P_D$, and confirm that if networks grow with duplication, the scale-free edge distribution is unavoidable.

\begin{figure}[htbp] 
   \centering
   \includegraphics[width=3in]{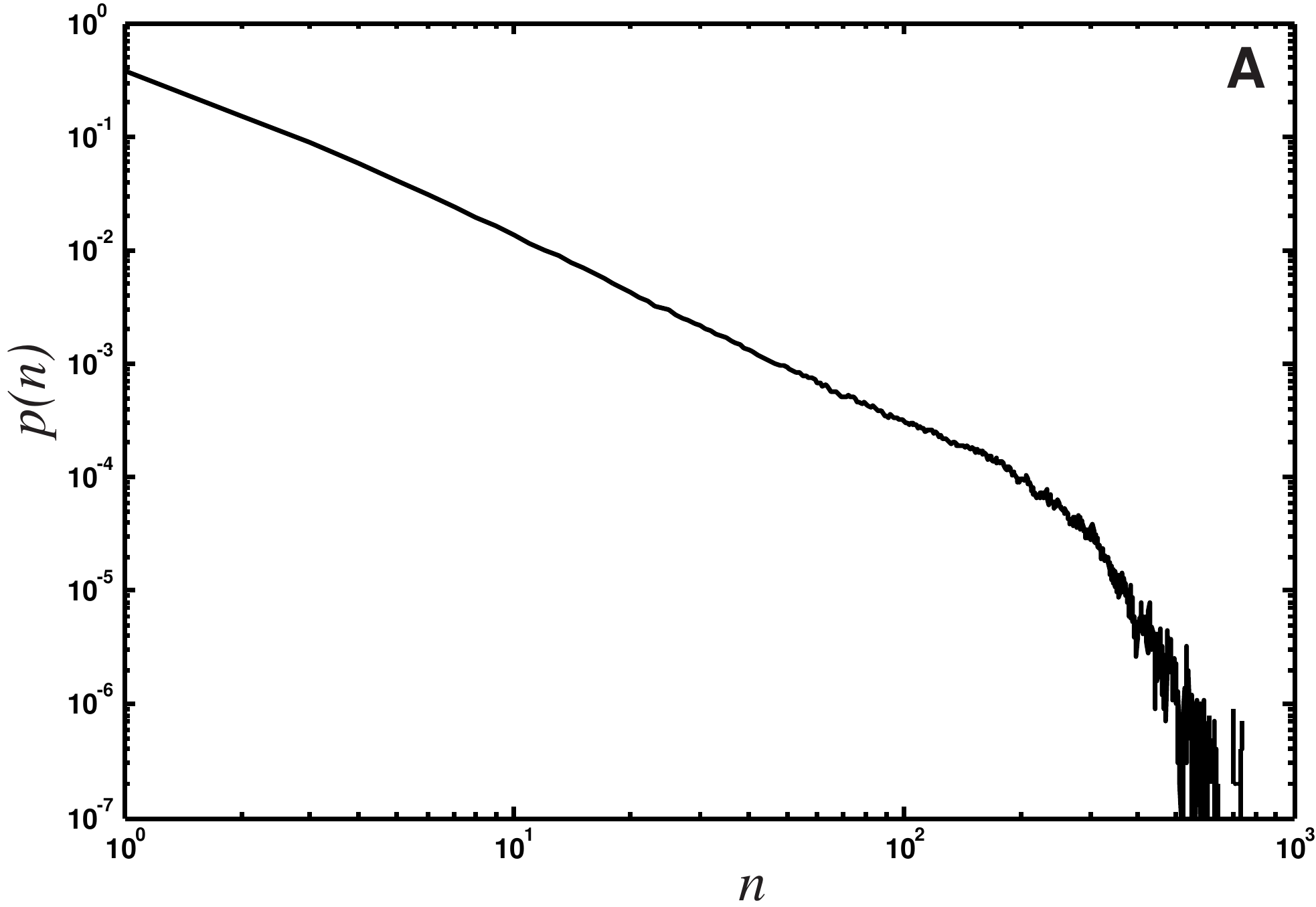}
      \includegraphics[width=3in]{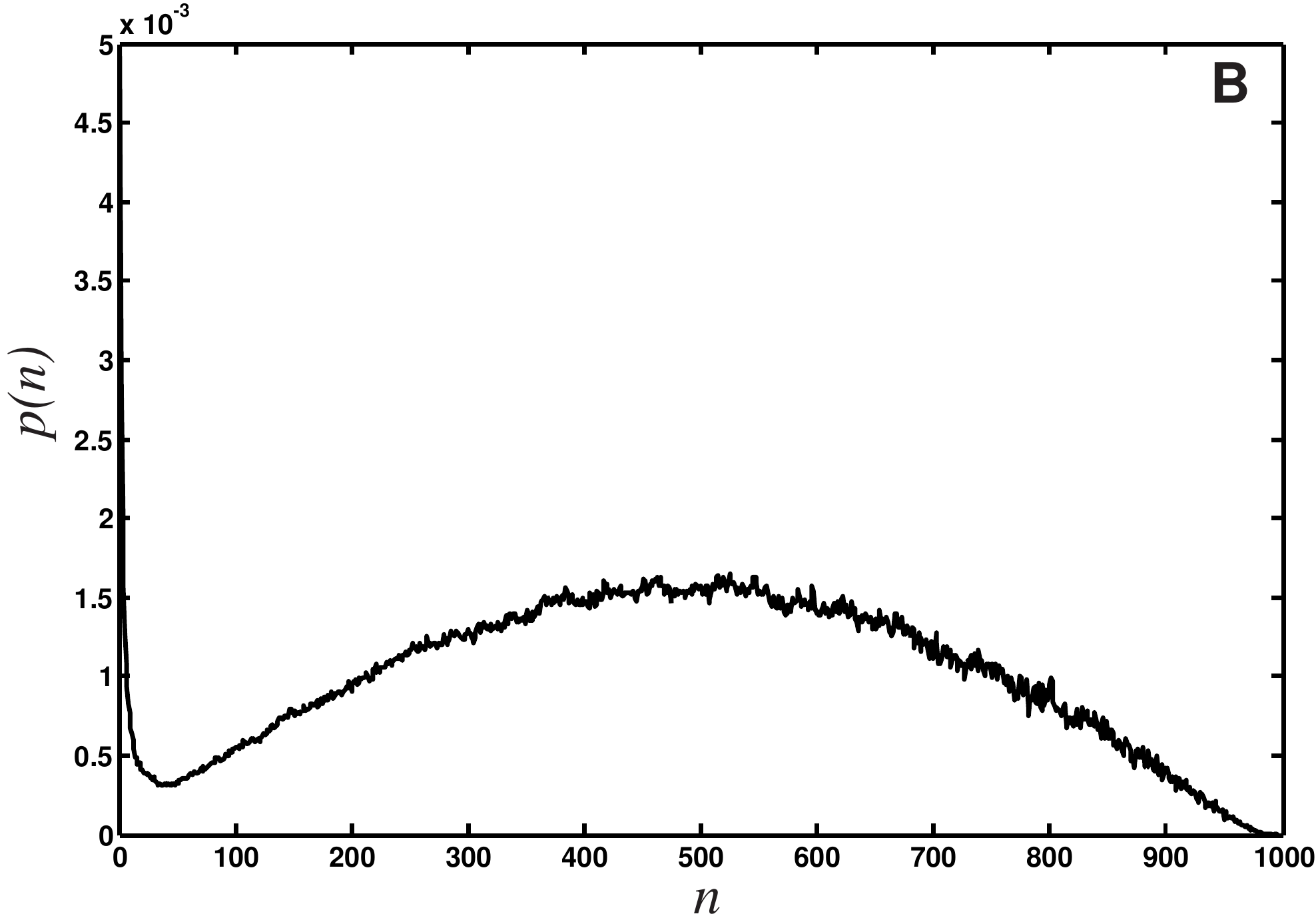}
   \caption{{\bf Edge distribution of networks grown with different parameters} (A) Scale-free edge distribution of networks obtained with a growth algorithm with $P_N=0.2625$, $p=1.0$, $P_E=0.15$, $q=1.0$, $P_D=0.225$, and $r=1.0$, undirected edges, no modules, average over 1,000 networks grown to $10,000$ nodes. (B) Edge distribution of networks grown with $P_N=0.002625$, all other parameters as in (A), averaged over 10,000 networks grown to 1,000 nodes.}
   \label{fig:dist}
\end{figure}

The most ``pure" scale-free networks actually emerge if nodes are added more often than edges (see Fig.~\ref{fig:dist}A). Choosing a low $P_N$, on the other hand, leads to the growth of networks with a Poissonian degree distribution (Fig.~\ref{fig:dist}B). Note that Figs.~\ref{fig:dist} were obtained with the same set of parameters except that the node addition probability was 100 times less for the graph that resulted in an \er-like edge distribution. In principle, keeping the relative ratio of the three probabilities $P_N$, $P_E$, and $P_D$ the same (when $p=q=r=1$) results in the same edge distribution (see {\bf Models}).

\begin{figure}[htbp] 
   \centering
   \includegraphics[width=3.5in]{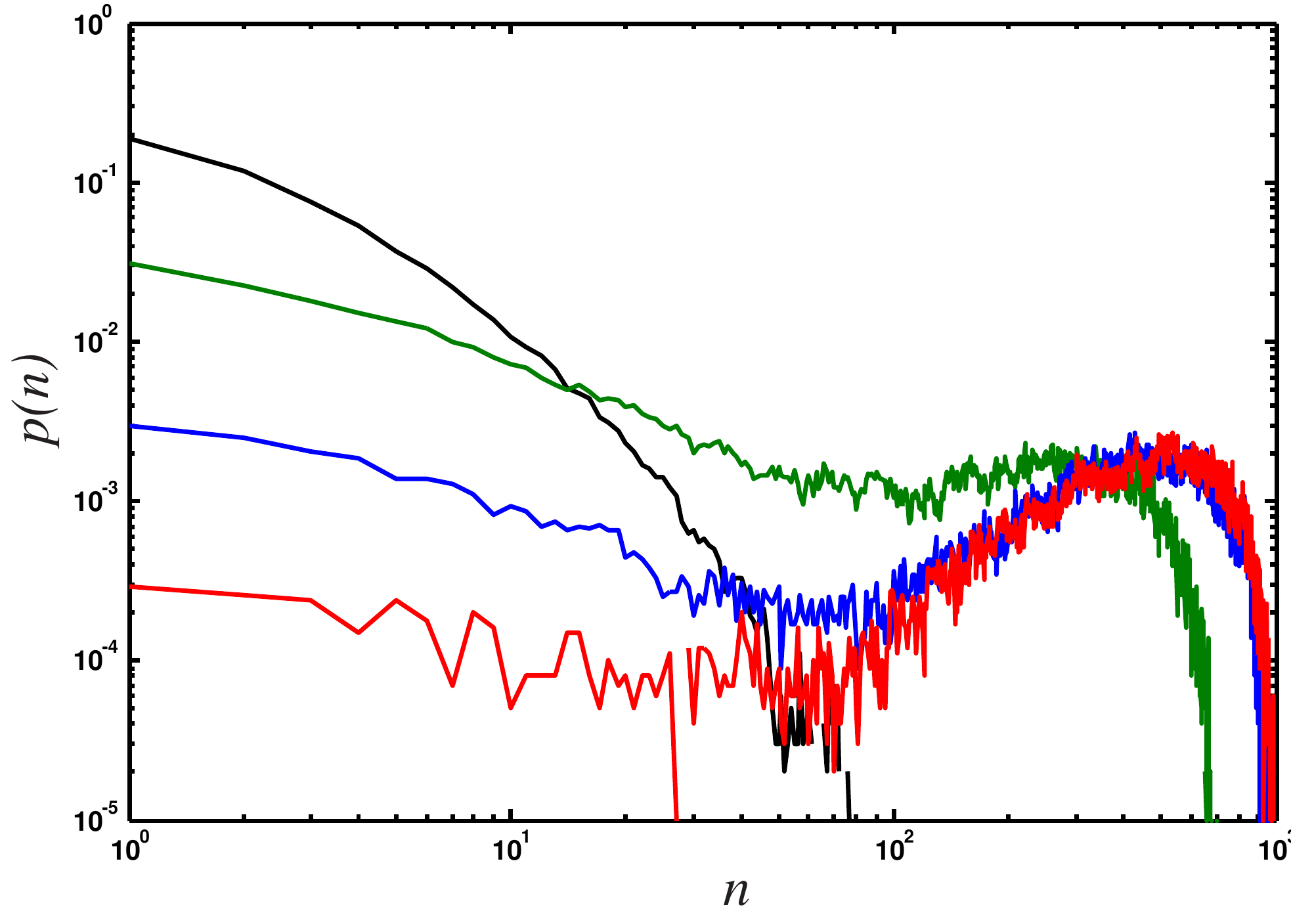}
   \caption{{\bf Edge distributions for networks grown under different regimes.} $P_N=1.0$ (black line, exponential distribution) $P_N=0.1$ (green), $P_N=0.01$ (blue), and $P_N=0.001$ (red). All other parameters are set to $P_E=p=q=0.75$, $P_D= 0.5, r=1.0$. Networks are unmodular and undirected, grown to size $n=10,000$, averaged over 100 replicates.}
   \label{fig:trans}
\end{figure}

Choosing probabilities in between the parameter values described allows us to grow any network, with edge distributions in between exponential and Poisson. For example, there are interesting ``transition stages" where parameter combinations lead to networks that are neither scale-free nor \er. We show in Fig.~\ref{fig:trans} extreme and intermediate edge distributions where we varied the node event probability ($P_N$) from 0.001 to 1.0 while keeping all other probabilities constant. The distribution obtained for $P_N=0.001$ 
has all the characteristics of an \er-type edge distribution, such as the one depicted in Fig.~\ref{fig:dist}B (note the difference in scales). We conclude that the edge distribution can be controlled entirely with the node addition probability and the edge duplication probability (as long as the edge addition probability is not too low): for low edge duplication, tuning $P_N$ from 1 to small values morphs the degree distribution from exponential to \er. If the edge duplication probability is substantial, however, the same change in $P_N$ moves the distribution from scale-free to \er. As a corollary, moving $P_D$ from small values to larger values for moderate to high $P_N$ changes an exponential towards a scale-free distribution, as we saw in Fig.~\ref{fig:degdist}.  

The algorithm can be used to create lattices with an arbitrary degree or connectivity by making use of the assortativity matrix in an unconventional manner: Each node of the lattice is assigned a unique module, where the probability of having an edge between modules reflects the desired neighborhood relations in the lattice. Instead of seeding the growth process with a single node, the algorithm is started with a fixed number of nodes and no edges, and $P_N=0$. The growth process then enacts a percolation problem with edge probability $qP_E$, and a geometry dictated by the assortativity matrix.

\begin{figure}[htbp] 
   \centering
   \includegraphics[width=4in]{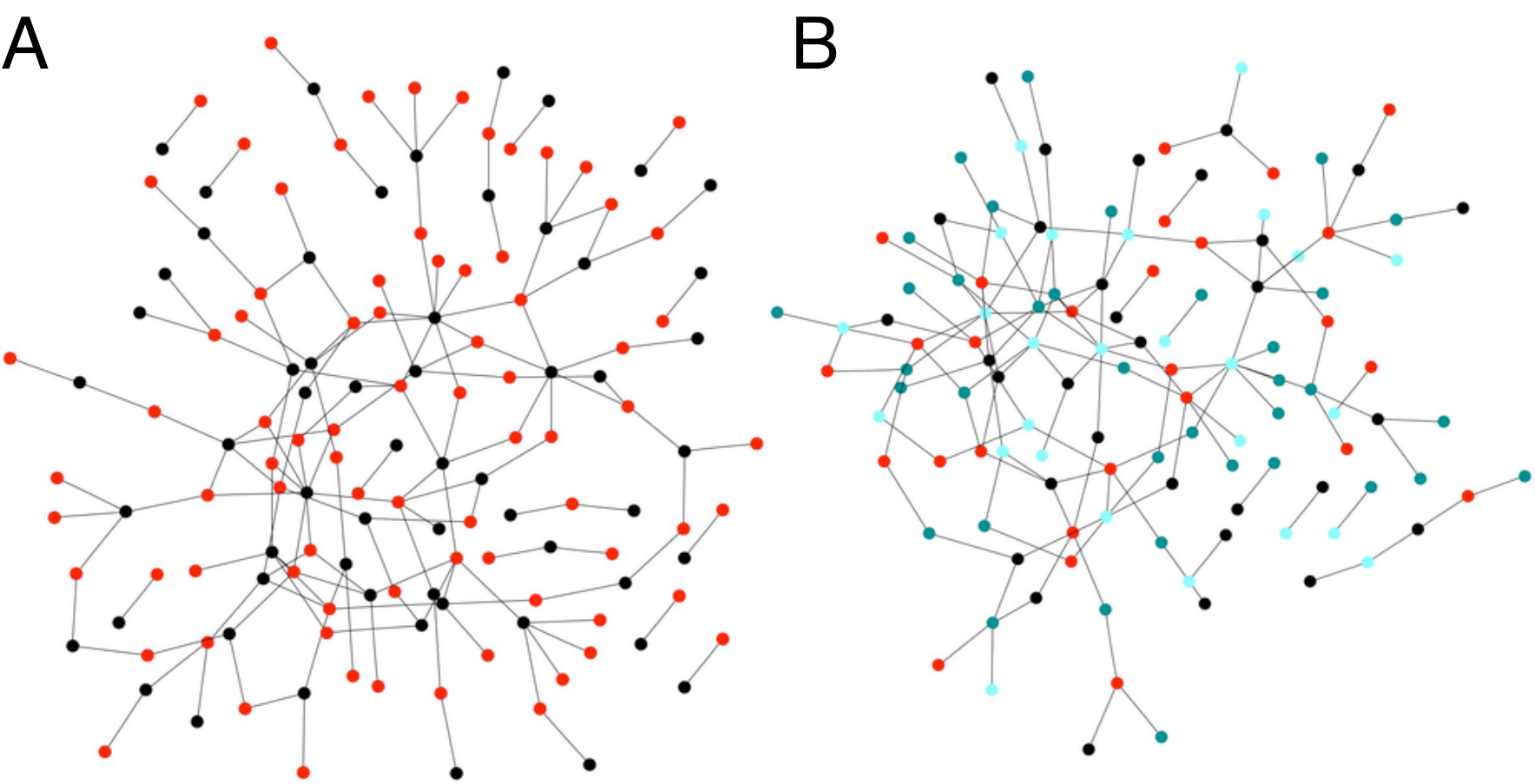}
   \caption{{\bf Bipartite and $k$-partite graphs.} (A) Bipartite graph, nodes from one group colored in red, nodes from the other group colored in black, $P_N/P_E$=0.3, $p=0.85$, $q=0.75$, $P_D=0.0$. The graph was grown for 1,000 iterations of the algorithm, with undirected edges. (B) Graph grown with the same parameters as (A), but for $k=4$. }
   \label{fig:bipart}
\end{figure}

To create bipartite graphs with edges only connecting nodes from different groups, we can grow networks from an assortativity matrix with a vanishing diagonal  [see Fig.~\ref{fig:bipart}A]. Nearly bipartite graphs are obtained by varying the entries in the matrix accordingly.

Clearly, the algorithm can generate arbitrary $k$-partite graphs, by extending the dimension of the assortativity matrix. We show in Figure \ref{fig:bipart}B a network iterated for 1,000 steps with the same parameters as Fig.~\ref{fig:bipart}A, but with $k=4$ (nodes colored according to the group label).

\subsection*{Modularity}
That biological, technological, and social networks are organized in a modular fashion is by now a commonplace observation. Yet, there is no standard measure of modularity, nor is there a standard algorithm that will partition networks into modules. There are several reasons for this apparent shortcoming. On the one hand, while the term ``modular organization" is fairly intuitive, anyone who is familiar with the structure of real-world networks understands that this intuitive notion can only be applied approximately, and with a good amount of prudence. Modules are often identified using the topological structure of the network, for example by counting the number of shortest paths between nodes, or by identifying an excess number of edges between nodes as compared to a random network. However, it is also possible that groups of nodes function together as a module without any obvious topological signature. Furthermore, functional modules often overlap, while topological modules are usually defined in such a way that they are mutually exclusive. Therefore, we expect that topological and functional measures of network modularity can disagree, and that this disagreement can be more or less severe depending on the type of network under consideration. 

We would also like to highlight the difference between modularity {\em measures}, which quantify the modularity in a network whose modules have already been determined, and {\em module-discovery} algorithms, which partition a network into groups of nodes. Often, module-discovery is performed by attempting to maximize a modularity measure, but in principle neither does a modularity measure imply an algorithm for module discovery, nor does a module-discovery algorithm necessitate a measure of modularity. 

A commonly used measure of modularity is due to Newman~\cite{Newman2006}, who assumes that
modularity implies that nodes that are in the same module have more connections between them than would be expected for a random network, that is, a network where all the module assignments have been randomized. If $k_i$ is the number of edges of node $i$, and $m=1/2\sum_i k_i$ is the total number of edges in the network, then the probability that two nodes $i$ and $j$ are connected by chance is $k_i k_j/2m$ (as long as the degrees for node $i$ and $j$ are independent). Now, define the network adjacency matrix $A$, in such a way that $A_{ij}=1$ if node $i$ connects to node $j$ and $A_{ij}=0$ otherwise. This matrix is symmetric for undirected networks, and can have non-integer entries if the {\em strength} of a connection is taken into  account. Here, we limit ourselves to undirected networks that have ``binary" edges, but the extension is obvious. We furthermore limit ourselves to networks without node self-connections, which implies that the diagonal of $A$ vanishes. If furthermore the module {\em assignment} for each node is known, we can define a modularity matrix $S$ in such a way that $S_{ij}=1$ if nodes $i$ and $j$ are in the same module, and zero otherwise. Newman's modularity $Q_N$ is then defined as~\cite{Newman2006}
\be
Q_N=\frac1{2m}\sum_{i,j}\left(A_{ij}-\frac{k_ik_j}{2m}\right)S_{ij}\;. \label{QN}
\ee
There is clearly a certain amount of arbitrariness in modularity measures of this kind. For example, a different measure using similar ideas is often called the ``assortativity" of a network. This measure is also due to Newman~\cite{Newman2003b}, and quantifies how likely it is that nodes of the same ``kind" attach to each other, where ``kind" can be any tag that is attached to a node to distinguish it from another class of nodes. In the following, we refer to this tag as the node's {\em  color}, so that assortativity measures how often nodes of the same color connect to each other rather than to nodes of a different color. Let us define the assortativity matrix $e$ (sometimes called the ``mixing matrix") such that $e_{k\ell}$ gives the fraction of edges that attach a node of color $k$ to a node of color $\ell$, and $a_k=\sum_\ell e_{kl}$ is the fraction of edges that either begin or end at a node of color $k$ (we again restrict ourselves to undirected networks here, so that $e$ is symmetric). Newman's assortativity is then given by~\cite{Newman2003b}
\be r=\frac{{\rm Tr}\, e -\sum_k a_k^2}{1-\sum_k a_k^2}\;. \label{assort}
\ee
Both measures (\ref{QN}) and (\ref{assort}) are bounded from above by 1, and they can both become negative (indeed, Newman's modularity and assortativity measures are closely related, see Supplementary Text). While the assortativity is constructed in such a way that networks with random assignment of colors (modules) to nodes gives rise to a vanishing measure, this is not generally true for $Q_N$. Furthermore, both measures can in principle detect in networks a tendency of nodes of the same module or color {\em not} to connect to each other (a phenomenon we call {\em anti-modularity} or {\em anti-assortativity}). However, the measures do not treat anti-modularity (or anti-assortativity) on the same footing as modularity or assortativity. 
It is possible to introduce a measure of modularity that is closely related to both of Newman's measures, but gives more weight to ``like"-edges if the number of colors is large. This is obtained by modifying the modularity matrix that enters Eq.~(\ref{QN}) so that
\be
\tilde S_{ij}=\left\{\begin{array}{l}1\ \ \ \ {\rm if}\  i\ {\rm in\ same\ module\ as}\ j\ \\
-\frac{1}{N_c-1}\ \ \ {\rm otherwise}
\end{array}\right.\;,
\ee
where $N_c$ is the number of modules or colors. (As in the following we will tag nodes that belong to the same {\em functional} module with the same color, we often refer to colors or modules interchangeably.) With such a modularity matrix, connections between nodes of unlike color are penalized, most heavily so if there are only a few colors. We define our functional modularity measure in terms of this generalized modularity matrix
\be
Q_H=\frac1{2m}\sum_{i,j}A_{ij}\tilde S_{ij}\;, \label{QH}
\ee
but note that we omitted the term $-k_i k_j/2m$ in Newman's measure that subtracts the probability that two nodes connect at random. Indeed, the latter bias is typical for modularity measures that attempt to capture the way modules are reflected in network topology, while our measure $Q_H$ focuses on function only. Because $Q_H$ can also be written as (see Supplementary Text)
\be Q_H=\frac{N_c}{N_c-1}\left({\rm Tr}\, e-\frac1N_c\right)\;, \label{QHandTre}
\ee
we see that $Q_H$ vanishes for non-associative (non-modular) networks, because when  color is assigned randomly to nodes we have $e_{ij}=1/N_c^2$ so that ${\rm Tr}\,e=1/N_c$. 
At the same time, $Q_H$ is maximal for graphs if colors only connect to like colors (${\rm Tr}\,  e=1)$. But in contrast to Newman's measures, $Q_H$ can become significantly negative, more so if the number of modules is small. For bipartite graphs, for example (two colors where only unlike colors connect) we find ${\rm Tr}\,e=0$, so that $Q_H=-1$.

\begin{figure}[htbp] 
   \centering
   \includegraphics[width=3.2in]{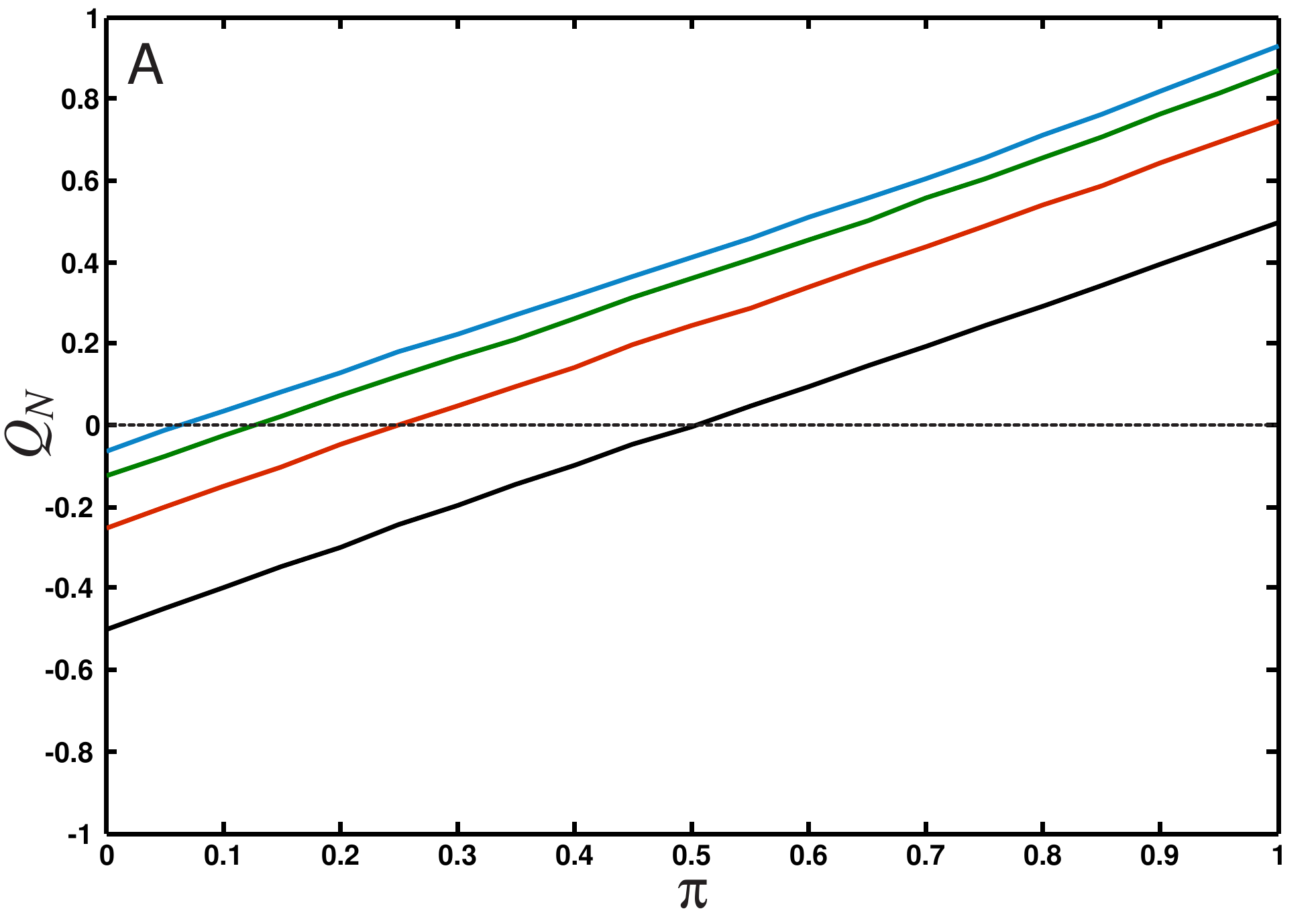}   \includegraphics[width=3.2in]{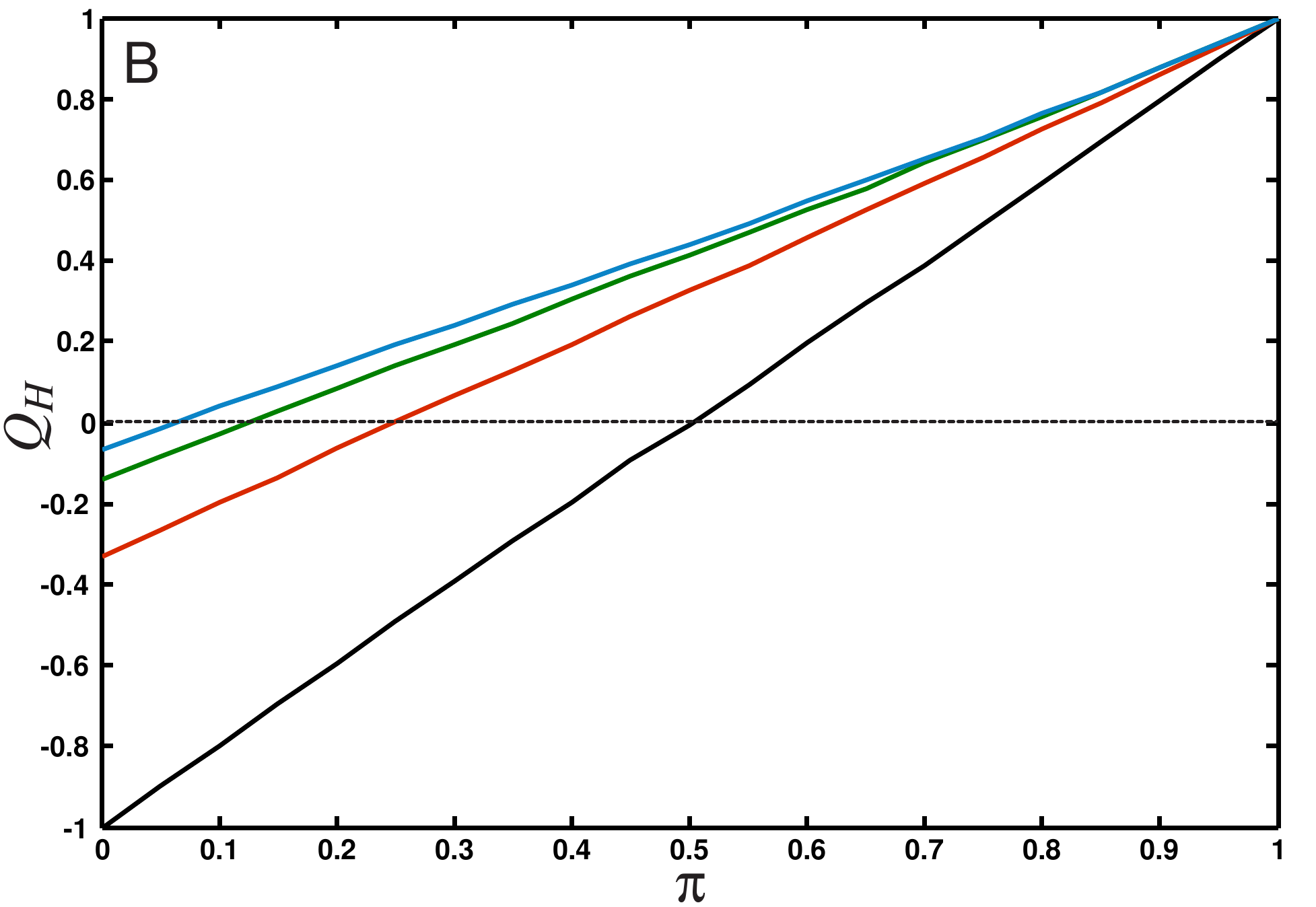}
   \caption{{\bf Comparison of modularity metrics} Comparison of the modularity measures defined in Eqs.~(\ref{QN}) and (\ref{QH}). Networks with between 2 and 16 modules were grown depending on the intra-module edge probability $\pi$. (For $\pi=0$ the networks are as anti-modular as possible and become $k$-partite (where $k$ is the number of modules), while for $\pi=1$ they are as modular as possible). (A): $Q_N$ [defined in Eq.~(\ref{QN})] for $N_c=2$ (2 modules, black line), $N_c=4$ (red), $N_c=8$ (green), $N_c=16$ (blue).  (B) $Q_H$ [defined in Eq.~(\ref{QH})]. Colors as in (A).  Each point was averaged over 50 networks with 1,000 nodes. The networks were grown with $P_N=0.5, p=1.0, P_E=1.0, q=1.0$ and  $P_D=0.0$, using undirected edges.}
   \label{fig:antimod}
\end{figure}

To study how the different modularity measures depend on the number of modules in the network as well as the strength of the module's interconnectivity, we generate networks with a tunable amount of modularity. A simple model for generating modular networks is an assortativity matrix for $N_c$ colors where like-colors connect to each other with probability $\pi$ (the intra-module edge probability), and connect to nodes of a different color with probability $(1-\pi)/(N_c-1)$, irrespective of color (the ``equal opportunity" model, see Supplementary Text). The probability $1-\pi$ can then be viewed as an {\em inter-module} edge probability and can be used to dial between perfectly modular ($\pi=1$) and perfectly anti-modular $(\pi=0)$ networks. 
The functional modularity $Q_H$ seen in Fig.~\ref{fig:antimod}B depends strongly on the number of modules, and is larger than $Q_N$ (depicted in Fig.~\ref{fig:antimod}A) for modular networks, and smaller than $Q_N$ for the anti-modular ones. Indeed, the inherent bias in Newman's measure for modules whose member nodes are strongly connected to each other leads to an underestimate of the modularity
for strongly connected modular graphs, and an equally underestimated antimodularity for multipartite graphs, as compared to the measure $Q_H$, when the number of modules is small. 

The functional measure $Q_H$, in turn, cannot be used to {\em detect} the number of modules or communities, for precisely this reason: because no connection bias is assumed, there are no topological means to identify clusters. If the number of modules is given, on the other hand, $Q_H$ can be used to guide a graph partitioning algorithm. Note that the measures become indistinguishable in the limit of an infinite number of modules. 

\begin{figure}[bhtbp] 
   \centering
   \includegraphics[width=3.75in]{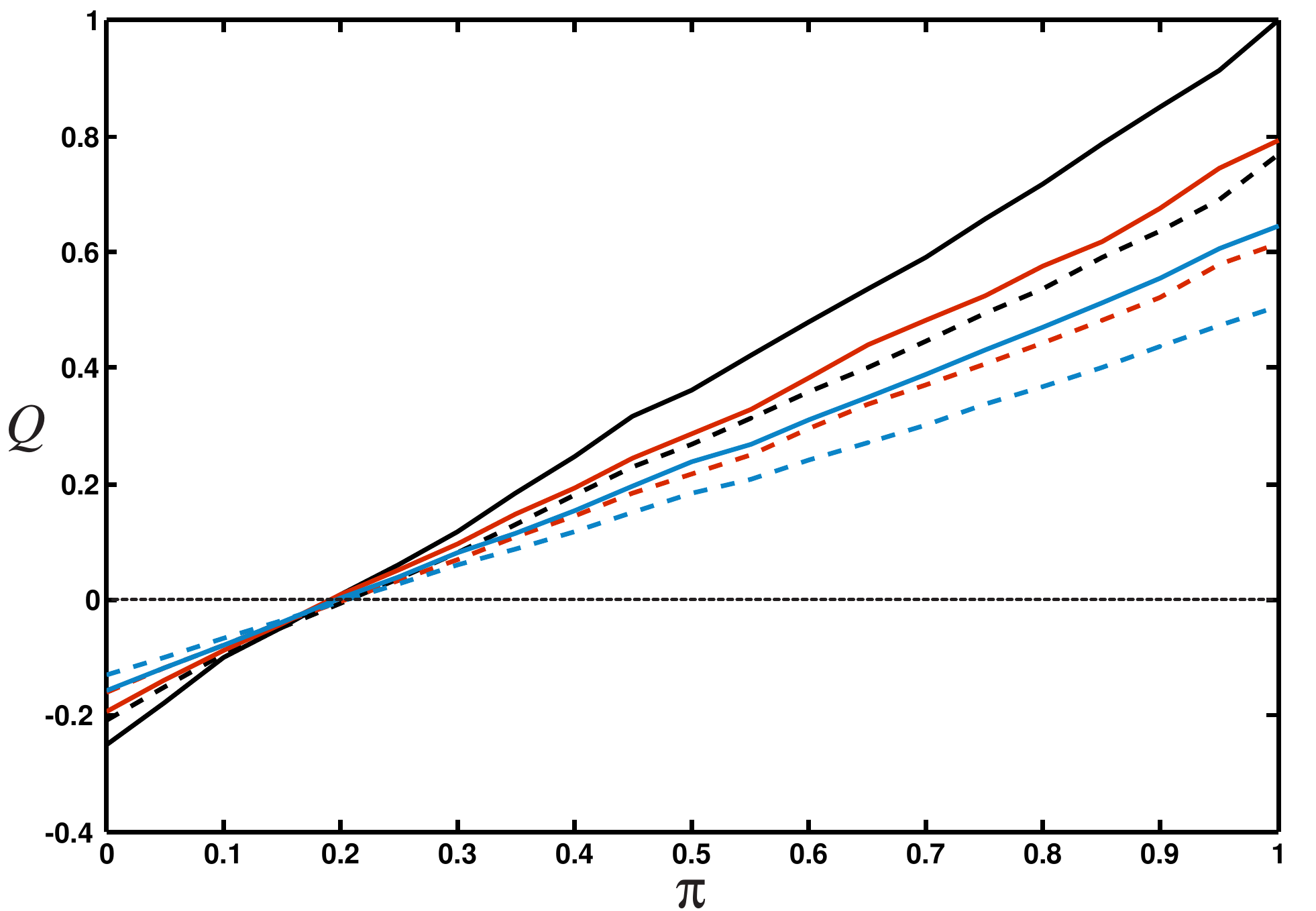}  
   \caption{{\bf Modularity depends on node-fusion probability}. Average modularity (from 50 independent networks with five modules, grown to 1000 nodes) for different node duplication probabilities $r=0.5$ (blue), $r=0.75$ (red), and  $r=1.0$ (black), for the modularity measures $Q_N$ (solid lines) and $Q_H$ (dashed lines). The networks were grown with the stochastic parameters set to 
   $P_N=0.5, P_E =1$, $p=q=1.0$, $P_D=0.2$.}
   \label{fig:mod}
\end{figure}

We can also investigate the impact node {\em fusion} has on modularity (Figure \ref{fig:mod}), for the modularity measures $Q_N$ and $Q_H$, by studying how modularity depends on module strength in networks grown with different node fusion probabilities.
Irrespective of the measure, modularity is highest if nodes are not fused ($r=1$) and decreases as the node fusion probability increases ($r<1$) because node fusion is blind to the module assignment, while node duplication creates another node with the same color and the same edges as the original node.  The larger the probability for adding an edge within modules is (larger $\pi$), the more modular the networks are, as expected. Because $Q_H$ does not penalize modules if they do not have an excess of edges between them, $Q_H$ is mostly larger than $Q_N$. For small $\pi$, more connections exist between nodes of {\em different modules} than within them, so that both modularity measures become {\em negative}.

The impact of node {\em duplication} on modularity more complicated. On the one hand, because node duplication brings with it the duplication of the edges that the duplicated node is attached to, whether or not node duplication leads to an increase in modularity depends on whether the network sports more inter-module or more intra-module edges. On the other hand, node duplication can skew the fraction of nodes that belong to any particular module by amplifying stochastic events that occur early-on in network growth. While node colors are chosen either randomly or according to a node probability vector when a node is created (see {\bf Model}), the color of a node (that is, its module membership) is inherited under duplication. As a consequence, module {\em sizes} fluctuate considerably across different realizations of the network, and the modularity can become significantly different from that predicted by the $e$-matrix generating the network. A detailed analysis of duplication on modularity is beyond the scope of this manuscript, and will be presented elsewhere.
\subsection*{Global properties}
A number of interesting global topological properties have been observed in networks, both in the case of biological or engineering networks that are built via growth processes, and in random networks that form via random edge addition. Foremost in the first category is the ``small-world" effect: the observation that many biological and technological networks have a short mean path between nodes (as compared to an equivalent randomized network), while being highly clustered (again with respect to a randomized equivalent network~\cite{WattsStrogatz1998}, see also the review~\cite{HumphriesGurney2008}). 
Humphries and Gurney~\cite{HumphriesGurney2008} introduced a quantitative measure to study the ``small-world-ness" of a network, which is particularly useful because networks that have a high edge-density can automatically appear to be in the small world class, but trivially so. Following Humphries and Gurney, we define the ratio of the ``mean shortest path between nodes" in a network to the mean shortest path in the randomized version of the network (an \er\ network with the same number of nodes and edges):
\be
\lambda_g=\frac {L_g}{L_{\rm random}}\;,
\ee
and the ratio of the graph clustering coefficient $C_g^{\Delta}$ with respect to that of the randomized version $C_{\rm random}^\Delta$
\be
\gamma_g^\Delta=\frac{C_g^{\Delta}}{C_{\rm random}^\Delta}\;.
\ee
The symbol $\Delta$ in the superscript of the clustering coefficients serves to remind us that this coefficient is obtained by counting the number of ``triangles" of nodes normalized to the number of pairs~\cite{Newmanetal2000}, which can be different from the clustering coefficient obtained by averaging the number of edges of the adjacent nodes~\cite{WattsStrogatz1998}. 

As small-world networks are identified by having a large $\gamma_g$ and a small $\lambda_g$, the ratio of these quantities can be used to measure small-world-ness:
\be
S^\Delta=\frac{\gamma_g^\Delta}{\lambda_g}\;. \label{sdelta}
\ee

\begin{figure}[htbp] 
   \centering
   \includegraphics[width=3.5in]{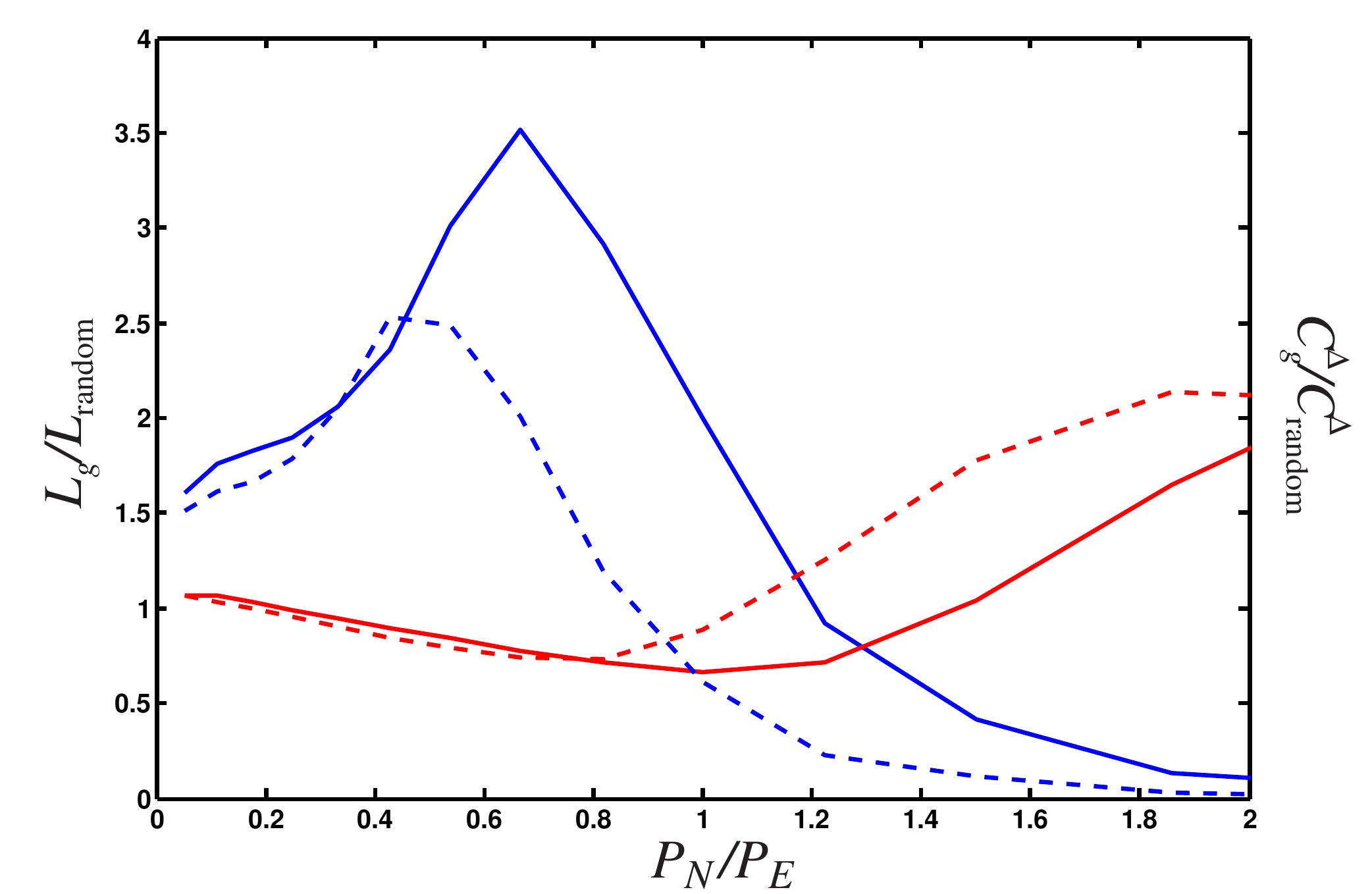}    
   \caption{{\bf Normalized mean shortest path and correlation coefficient.} The normalized mean shortest path between nodes, $L_g/L_{\rm random}$ (red lines) and the normalized correlation coefficient $C_g^\Delta/C_{\rm random}^\Delta$ (blue lines) as a function of the ratio $P_N/P_E$, for a node duplication probability $p=1$ and two different edge addition probabilities $q=0.75$ (dashed) and $q=1$ (solid). Average over 1,000 networks grown to 200 nodes, with $P_D=0$.}
   \label{fig:sw}
\end{figure}

We show the behavior of $\lambda_g$ and $\gamma_g^\Delta$ in Fig.~\ref{fig:sw} as a function of the ratio of node to edge event probability $P_N/P_E$ for networks grown to a size of 200 nodes.
As more and more nodes are added per edge-addition event (increasing ratio $P_N/P_E$ towards 1), the normalized mean shortest path first drops, and indeed, as long as $P_N/P_E<1.5$, the shortest paths in these networks are {\em shorter} than those in random networks. But once passed this threshold, the addition of more nodes without a commensurate increase in edges leads to longer and longer shortest paths (see Fig.~\ref{fig:sw}, solid red line). The normalized correlation coefficient increases rapidly with an increasing ratio $P_N/P_E$ up until $P_N\approx 0.65 P_E$, after which the ratio drops very fast. 

We also tested how decreasing the conditional edge-addition probability $q$ affects $\lambda_g$ and $\gamma_g^\Delta$. We expect that a decrease in $q$ will move the mean shortest path and the correlation coefficient towards their random graph equivalents, because a $q<1$ implies that sometimes edges are removed (for $q=1/2$ edges are added as often as they are removed during an edge event), and the edge removal/edge addition process is tantamount to a {\em randomization} of the graph. We do indeed observe that $C_g^\Delta\rightarrow C_{\rm random}$ as $q$ decreases (we show the case of $q=0.75$ in Fig.~\ref{fig:sw}),  but $L_g/L_{\rm random}$ actually {\em increases} for decreasing $q$ as long as $P_N/P_E\lesssim 2$. 

\begin{figure}[htbp] 
   \centering
   \includegraphics[width=3.5in]{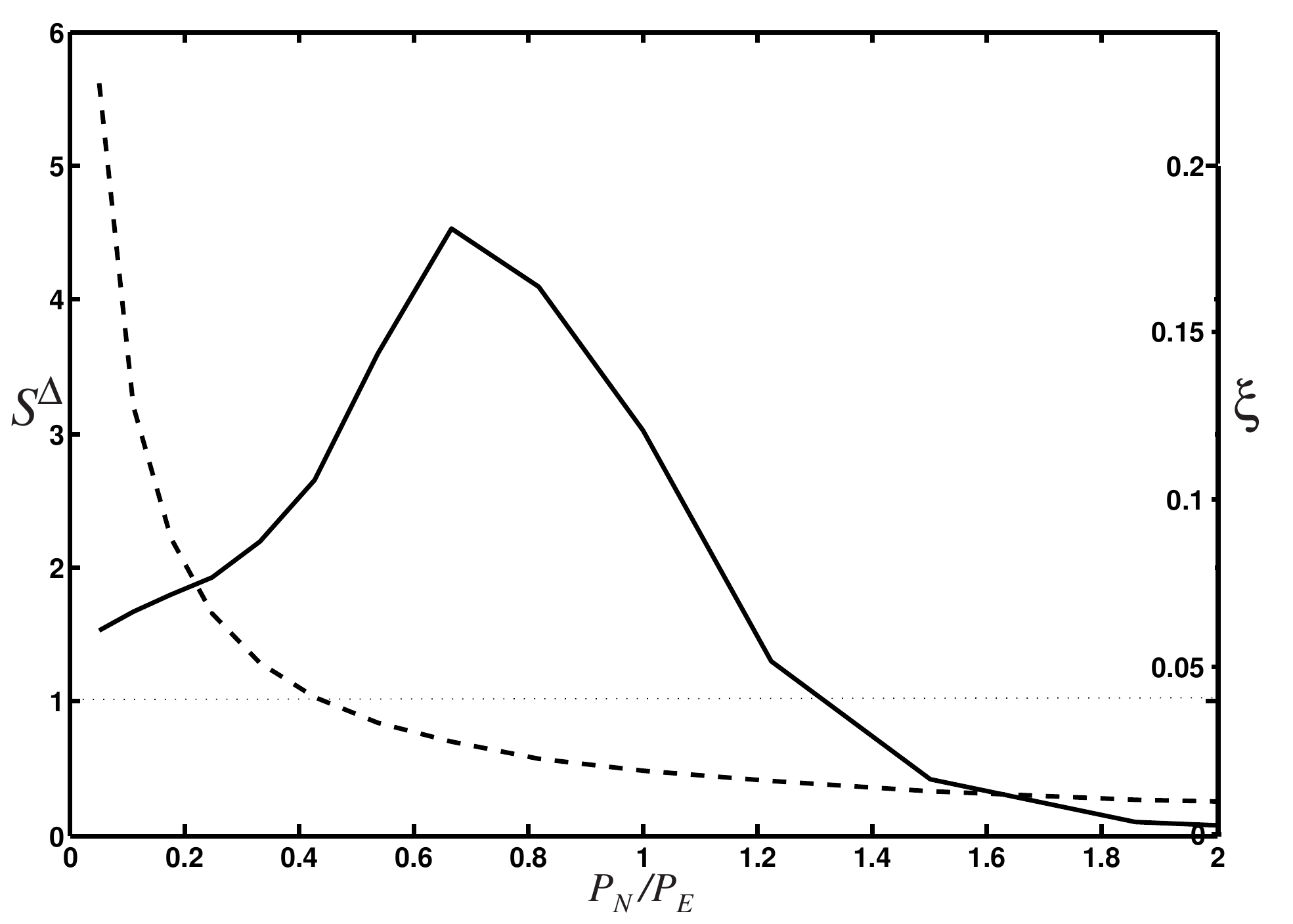}  
   \caption{{\bf Small-worldness $S^\Delta$ as a function of $P_N/P_E$.} The quantitative measure of small-worldness $S^\Delta$ (solid line) and the edge density $\xi$ [Eq.~\ref{xi}] (dashed line) as a function of the node to edge addition probability ratio. Networks are non-trivally ``small-world" if $S^\Delta>1$ while the edge density is low (e.g., $\xi<0.1$).} 
   \label{fig:sdelta}
\end{figure}

In order to determine what graph-growth parameters give rise to small-world networks, we plot the ratio $S^\Delta$ defined in Eq.~(\ref{sdelta}) (which is just the ratio of the two curves depicted in Fig.~\ref{fig:sw}) as a function of the ratio $P_N/P_E$ (Fig.~\ref{fig:sdelta}).  In this figure, we also plot the edge density (or ``sparseness") of the network (here $m$ is again the number of edges, and $n$ the number of nodes in the network)
\be
\xi=\frac{ 2m}{n(n-1)}\;, \label{xi}
\ee
because it is known that networks with a high density of edges can be trivially of the small-world kind~\cite{HumphriesGurney2008}. We see from Figs.~\ref{fig:sw} and \ref{fig:sdelta} that networks grown with a ratio $P_N/P_E<1.3$ have a small-world character, and furthermore that this character is maintained even for small ratios $P_N/P_E$ down to about $P_N/P_E\approx 0.2$, where the edge density is $\xi\sim0.1$. This behavior is similar to that observed in the Watts-Strogatz model~\cite{WattsStrogatz1998}, which becomes ``trivially small-world" in the limit of increasing randomness~\cite{HumphriesGurney2008}. 

Figure \ref{fig:sdelta} suggests that networks with small-world character are an automatic by-product of a stochastic growth process where the edge-event probability is of the order of the node-event probability or larger (here, $P_E/P_N\gtrsim0.75$), while the small-world character becomes trivial if $P_E$ is many times $P_N$. This is a plausible scenario for a number of biological networks, where it is much more likely to create a new edge (for example, an interaction between two proteins via a gain-of-function mutation, or a regulatory interaction) than it is to create a new node (the evolution of a new protein {\em de novo} or via lateral gene transfer). We can also see that this regime is easily achieved in social networks, as long as the creation of an interaction between nodes is more common than the addition of a new member to the social network. 

\subsection*{Critical Behavior}
\begin{figure}[htbp] 
   \centering
   \includegraphics[width=3.5in]{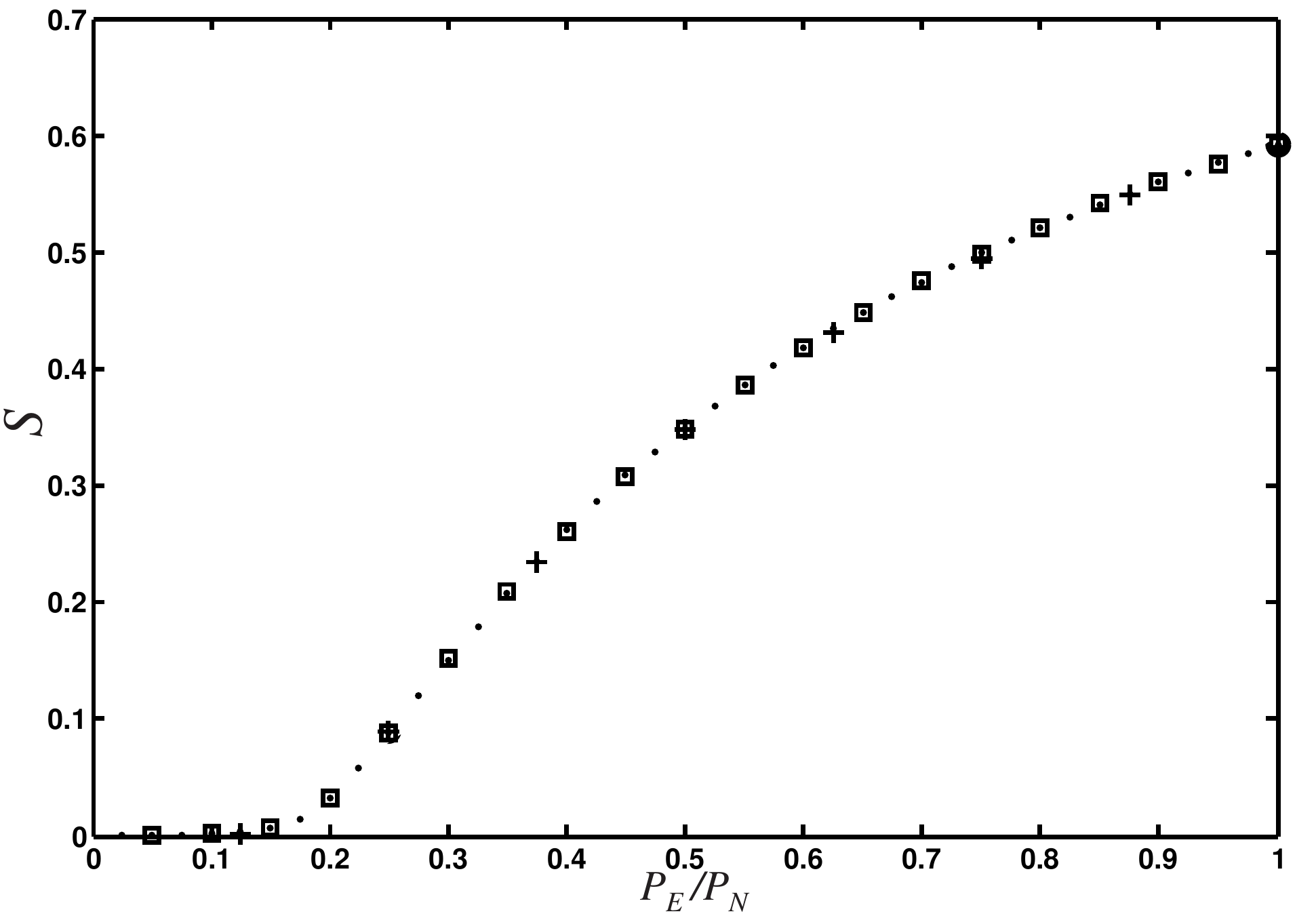} 
   \caption{{\bf Percolation phase transition in randomly grown networks.} Relative size of the giant component $S$ as a function of $P_E/P_N$, for networks grown with various combinations of $P_N$ and $P_E$: Crosses: $P_N=0.1$, circles: $P_N=0.2$, squares: $P_N=0.5$, dots: $P_N=1.0$ (with $P_D=0$ and $p=q=r=1$). Networks grown to 10,000 edges, average over 100 networks.}
   \label{fig:perc}
\end{figure}

Besides degree distribution, modularity, and small-worldness, a number of other global properties of networks have been studied in the literature that we can study with ease using our network growth model. It is known since the pioneering work of Erd\"os, Renyi, and Bollobas that static random graphs undergo a phase transition where a ``giant component" (a large connected component that scales with the size of the system) emerges at a critical probability of connecting edges. This phase transition is of the same kind as in percolation models, and is often referred to as the percolation transition in random graphs. Callaway et al.~\cite{Callawayetal2001},  Dorogovtsev et al.~\cite{Dorogovtsevetal2001}, as well as Bollobas et al.~\cite{Bollobasetal2005} have pointed out that random networks that are grown also undergo a percolation transition, but that this transition has a very different character: the critical point is infinitely differentiable (in fact, all the derivatives vanish at the critical point~\cite{Callawayetal2001,Dorogovtsevetal2001,Bollobasetal2005}). We can study this transition in our model as a function of the parameters not previously investigated, namely $P_N,p,q$, as well $P_D$ and $r$.  We observe that the percolation phase transition only depends on the ratio $P_E/P_N$ (see Fig.~\ref{fig:perc}), that is, the size of the giant component $S$ only depends on the {\em relative} rate at which nodes and edges are added. Even varying $q$ (allowing for edge removals) does not change this transition, as long as we plot the giant components versus $qP_E/P_N$ instead (results not shown). Note that this combination of parameters equals to the mean number of edges $\la k\ra$ (see {\bf Models}). 

\begin{figure}[htbp] 
   \centering
   \includegraphics[width=3.5in]{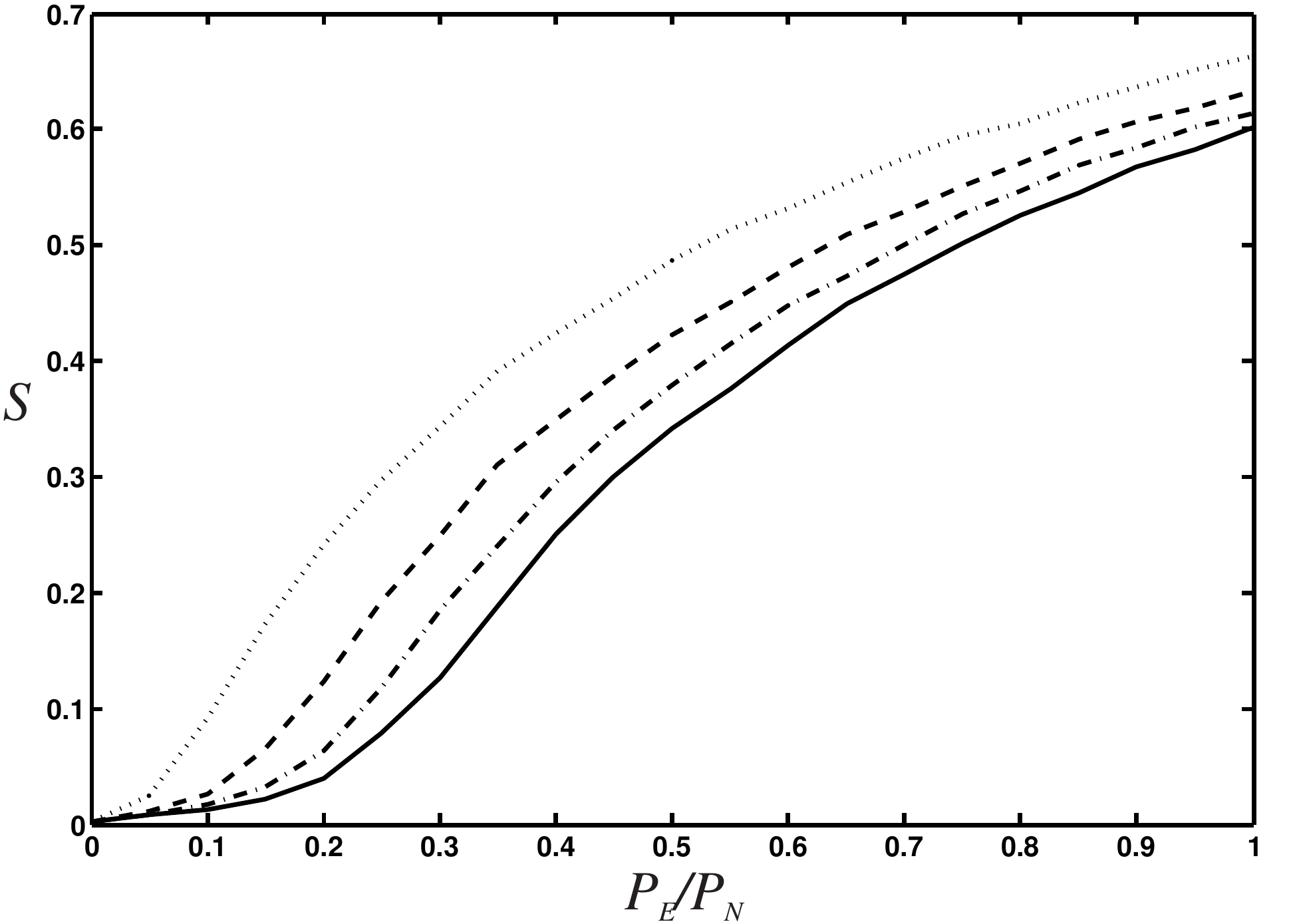} 
   \caption{{\bf Percolation phase transition with node fusion.} The relative size of the largest connected component $S$ as a function of the relative edge to node event probability (with $p=1,q=1$, and $r=0$), for different node fusion probabilities (as $r$ sets the node duplication probability, the fusion probability $(1-r)P_D$ is simply given by $P_D$). Solid line: $P_D=0$, dash-dotted: $P_D=0.1$, dashed: $P_D=0.25$, and dotted: $P_D=0.5$.  Average of 100 replicates of networks grown to size $n=$100.}
   \label{fig:fusion}
\end{figure}

Similarly, allowing for node duplication does not change the transition, as duplications may change the absolute size of the giant component, but do not affect its emergence. Node {\em fusion}, on the other hand, does affect the emergence of the giant component because fusions can lead to the merger of two separate clusters. We show in Fig.~\ref{fig:fusion} the relative size of the largest connected component of networks grown with different node fusion probabilities ($P_D=0.0,0.1,0.25,0.5$, $r=0$), where the value $P_D=0$ serves as the control (no node fusion). As expected, the onset of the transition is earlier when nodes can fuse, because node fusion does not change the giant component if the nodes are in the same cluster (except for diminishing its size by one), whereas whole clusters are fused if the nodes that are fused belong to different clusters. The nature of the phase transition (infinitely differentiable critical point) is unchanged.
Of course, as nodes are fused, the network grows more slowly, but the shape of the curve in Fig.~\ref{fig:fusion} cannot be recovered simply by scaling $P_N$ and $P_E$ taking into account the modified number of nodes and edges for each fusion event (data not shown). 

\subsection*{Biological relevance}
Given the variability of the networks that can be generated with this model, we may ask whether it is universal in the sense that {\em all} biological networks can be characterized by the set of parameters (six constants plus the $e$-matrix). While of course this is not a well-posed question, we tested whether networks can be grown to have an edge distribution that is similar to well-known biological reference examples, and whether the set of parameters giving rise to these networks allows us, by analogy,  to generate hypotheses about the process that generates them. Specifically, we grew networks to resemble the edge distribution of the neuronal network of the nematode {\it C. elegans}~\cite{Chenetal2006}, as well as a network similar to the protein-protein interaction network of yeast~\cite{Regulyetal2006}.
The best current data set for the {\it C. elegans} ``brain" includes 280 of the 302 neurons and their connections~\cite{Chenetal2006}. We 
binned the edge distribution from this data set and searched the parameter space of the model (five parameters, no modules, undirected edges) for sets that grow networks of 280 nodes with an edge distribution that minimizes the root mean square difference of the corresponding binned edge distribution. Note that because a graph's properties are unchanged if the relative ratio of the three event probabilities is maintained (as long as neither of them becomes too small), we kept the largest of the parameters (here $P_E$) fixed. We verified that a search with six independent parameters gives rise to the same set if rescaled to $P_E=1$.

\begin{figure}[htbp] 
   \centering
   \includegraphics[width=3.0in]{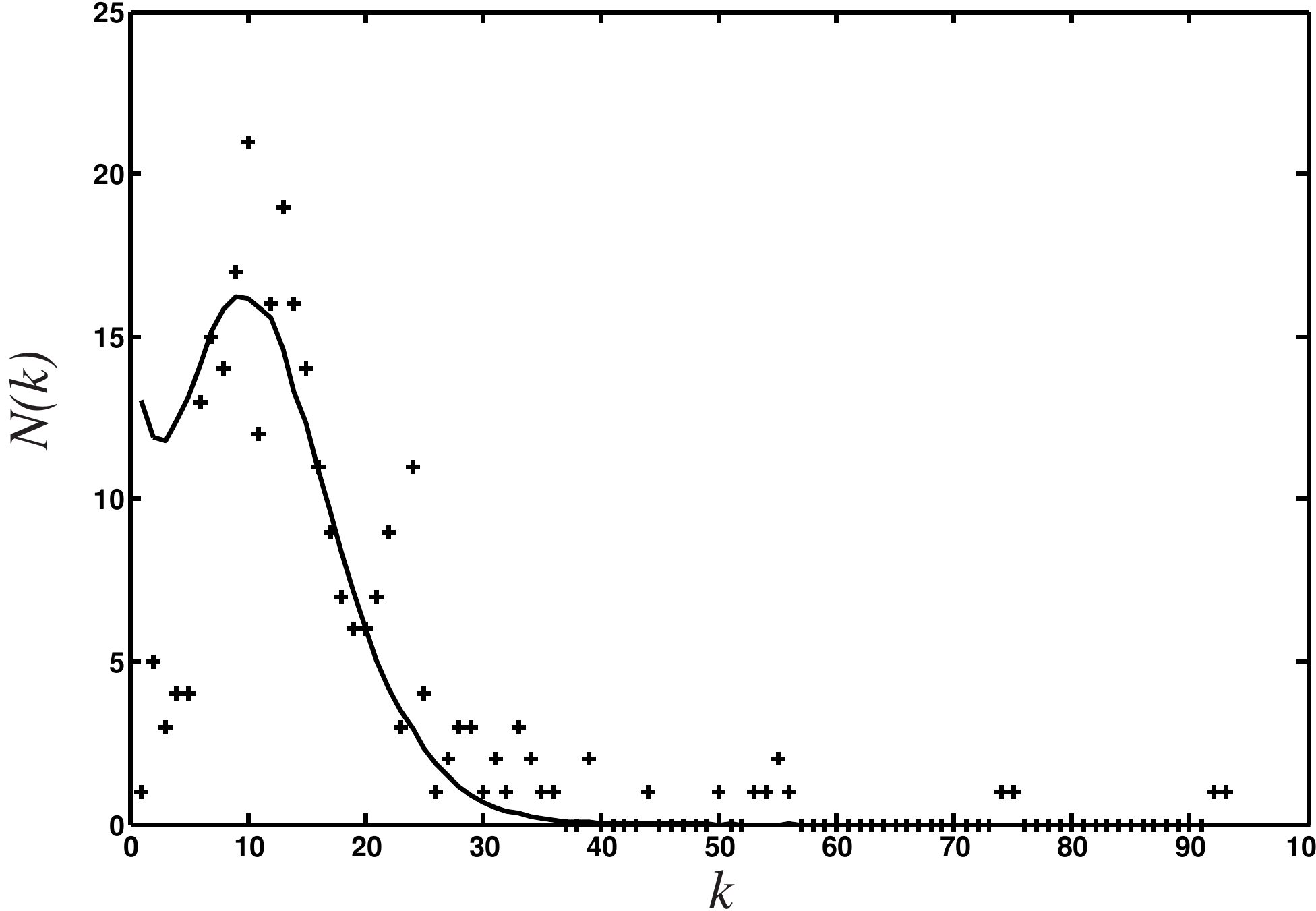}    \includegraphics[width=3.0in]{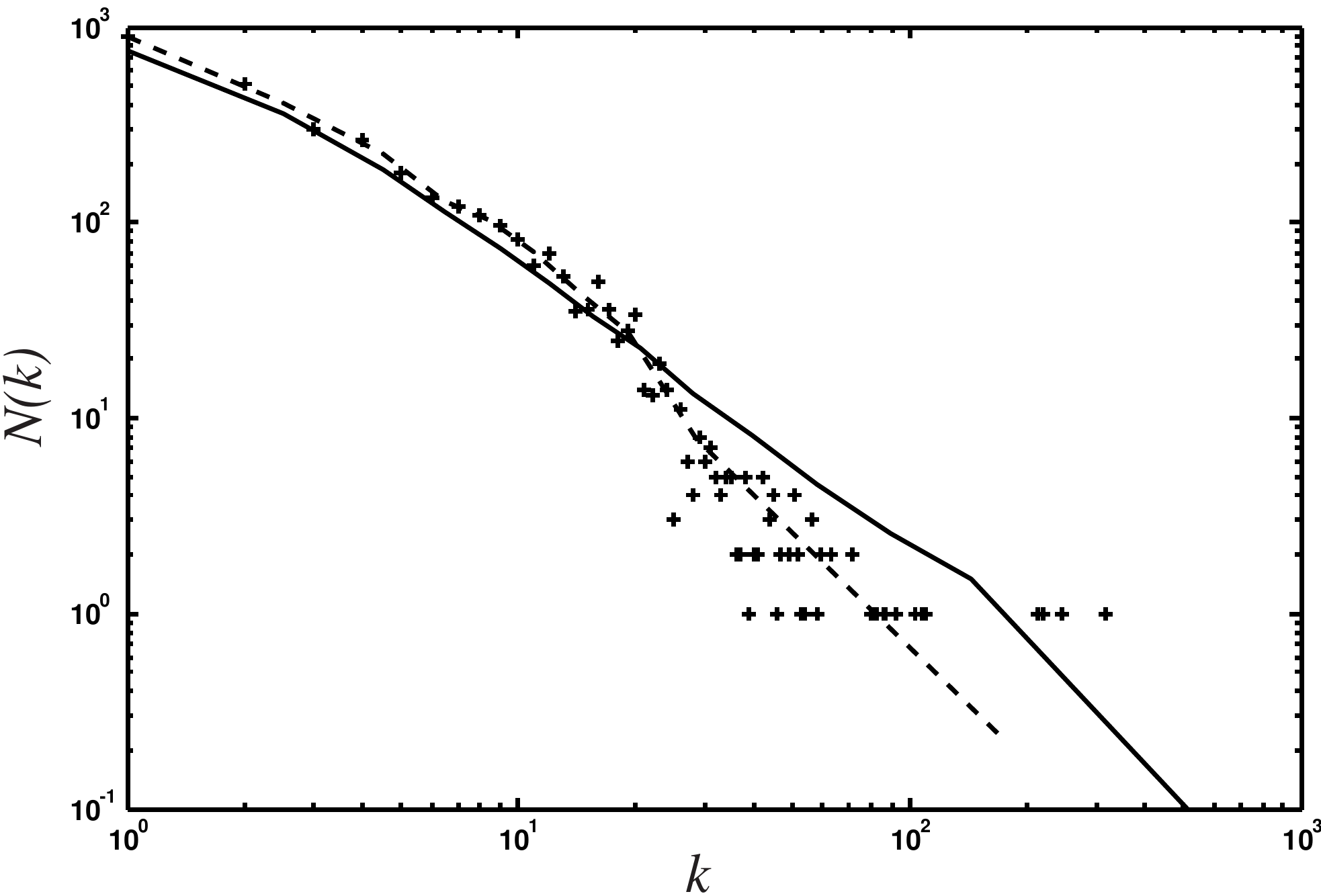} 
   \caption{{\bf Edge distributions of actual and simulated biological networks.} (A) Edge distribution of a network optimized to reproduce the edge distribution of the {\em C. elegans} neural network (solid line, circles), compared to the observed distribution (dashed line, diamonds). (B) Edge distribution of a network optimized to reproduce the edge distribution of the yeast protein-protein interaction network. The yeast data and the network growth data are binned using the threshold binning method~\protect\cite{AdamiChu2002}.}
   \label{fig:biol}
\end{figure}

Within the space of network parameters, the {\it C. elegans} network appears to be fairly rare, so that a straightforward Monte Carlo search often arrives at inferior fits. Our best solution is a network with $P_N=0.008$, $p= 0.71$, $P_E= 1.0$, $q=0.06$, $P_D= 0.028$, $r= 1.0$ ($P_E$ was fixed at 1.0 in this search). We show the degree distribution generated with this set of parameters (averaged over 1,000 realizations of the network) in Fig.~\ref{fig:biol}A (solid line). This set suggests that the {\it C. elegans} network reflects a growth process with a very small node addition probability, commensurate with our earlier observation that networks with an \er-like degree distributions are obtained using a small $P_N$. Such a small node addition probability is also consistent with the constraints imposed on {\em C. elegans} evolution by its invariant cell-lineage. The worm develops via {\em stereotyped cleavages} so that the patterns of cell division, differentiation, and death are the same from one individual to another: in the developing worm each cell has a predictable future, and each cell a well-defined set of neighbours~\cite{Sulstonetal1983}.  As a consequence, developmental changes giving rise to new nodes must be heavily constrained, as they would upset the delicate balance.  For the same reason, node duplications are also virtually absent in the simulated network. The small edge addition probability $q=0.06$ implies that edges are only added in 6\% of the edge events, and the network is consequently quite sparse. As the algorithm does not remove an edge if there is none between the randomly selected nodes, even such a small edge addition probability (in fact, it corresponds to a 0.94 edge {\em removal} probability)  always gives rise to an equilibrium edge count (calculated in {\bf Models} in the absence of duplication).

The simulation of the yeast protein-protein interaction network (interaction data from the highly curated set from Ref.~\cite{Regulyetal2006}) leads us to vastly different conclusions concerning the nature of the growth process. Because this network is much less rare, a Monte Carlo search converges fairly rapidly, and yields a similar set of parameters in all trials. For yeast, we started with 4 different initial conditions, conducted 5 trials each, and grew 20 replicas for each parameter set to obtain an average distribution, which we score by comparing the root mean square difference of the binned distribution to the binned yeast distribution.  We stop growth at 3,306 nodes (the size of the Reguly et al. network) and obtain a network that is remarkably similar to the yeast protein-protein interaction network, with $P_N=0.7\pm0.04$, $p=1.0\pm0.05$, $P_E=1.0\pm0$, $q=0.91\pm0.035$, $P_D=0.75\pm0.035$, $r=1.0\pm0.03$. Both the node addition probability $P_N$ and the node duplication probability $P_D$ are remarkably high. That the node duplication probability is this high for the generation of a protein-protein interaction data set is not surprising in the light of evidence that much of genomic evolution proceeds via gene duplication and subsequent diversification~\cite{Ohno1970,Zhang2003}.  However, the analysis also suggests that the yeast network edge distribution is only very approximately scale-free.

We conclude that particular degree distributions are (at least for the cases we examined) obtained with unique parameters sets, thus allowing us to entertain hypotheses about the processes that generated the networks we are simulating. Of course, such speculations rest on the assumption that other processes (such as for example, whole genome duplication, or horizontal transfer of sets of genes) do not play a role in shaping the network's edge distribution. Because we cannot rule out such processes in many of the standard networks, such a caveat always has to be issued.

\section*{Discussion}
We have presented an algorithm that, using only a few parameters, can generate networks of seemingly arbitrary degree distribution, modularity, and structure. Using this model, we were able to study how fundamental properties such as edge addition or removal, as well as node addition, removal, duplication, or fusion, affect a network's degree distribution, modularity, and structure. We found that we could grow networks with degree distributions anywhere between binomial, exponential, and scale-free, within a single framework or process. By introducing an assortativity matrix for the generation of nodes with different functional tags, we could furthermore grow networks with different degrees of modularity, by specifying the probability that a node of one ``color" will attach to a node with a different color. Once modules or functional groups have been identified in any real network, control networks can be grown that mimic the connection pattern or modularity of that network. One obvious example is again the {\em C. elegans} neuronal network. Its nodes can be divided into three classes: sensor neurons, motor neurons, and interneurons~\cite{Chenetal2006}. The $e$-matrix for this network can be reconstructed from the frequencies of inter-color edges, and be used to grow networks that not only have the same degree distribution as the original network, but mimic the connection pattern between functional types as well. 

We also introduced a modified modularity measure $Q_H$ for networks that is based entirely on the functional characterization of a network's nodes, rather than the connection pattern. This measure is neither better or worse than any of the existing modularity measures (such as Newman's $Q_N$ or the assortativity $r$), but rather highlights a different aspect of modularity. For example, while Newman's modularity $Q_N$ defines modules as those groups of nodes that are connected to each other more often than would be expected from the connection probability in  a random graph, the measure $Q_H$ does not assume such a bias. In fact, because most biological, social, and technological networks are not of the \er-type, it is often erroneous to compare the connection probability of modules to what would be expected in a random graph. This is particularly true for networks with scale-free edge distribution, which sport a number of hubs with many edges that do not necessarily connect to other nodes within the same module. The measure $Q_N$ will attempt to join such hubs in one and the same module, even if a measure based on betweenness centrality will separate them (see Supplementary Text). $Q_H$, in contrast, does not allow you to detect modules, but rather quantifies the modularity of a graph based  entirely on a previous group identification. Using such a measure, we can show that graphs are often less than modular, and can even become anti-modular. An extreme case of anti-modularity is given by bi-partite, and by extension, $k$-partite graphs. In fact, 
precisely modular and $k$-partite graphs appear as ``dual opposites" in this framework, obtained with an $e$-matrix with only ones on the diagonal and zeros elsewhere (divided by the number of modules), or else zeros on the diagonal for the $k$-partite graphs. 

The networks created using the present model recapitulate a large swath of existing literature concerning networks, of which we presented a selection here. For example, we were able to study how to generate networks with given global properties, such as small mean distance between nodes, or high clustering coefficient, and by extension examine the nature of small-world graphs. We were also able to study the percolation phase transition in networks, but unlike in the standard literature where the probability that edges are connected given a fixed set of nodes (or, as in Ref.~\cite{Callawayetal2001}, the node addition probability is fixed at $P_N=1$), we were able to study the size of the ``giant component" as a function of the ratio of the edge to node addition probability, and found the same phase transition. In addition, we could study the effect of node duplication and/or fusion on the nature and location of the transition.

Finally, we used the model to reverse engineer the growth parameters that might have led to the observed degree distributions of the {\em C. elegans} neuronal connection graph and the yeast protein-protein interaction network, while keeping in mind the assumptions behind that extrapolation. A Monte Carlo search process through the five-dimensional parameter space (not using modules) converged to suggest a unique set of parameters for each of those networks that led to biological conclusions compatible with our current knowledge about the forces that shaped these networks. We expect this model to be most useful in the generation of null models in the analysis of biological, technological, and social networks. The process easily generates networks with the same size and degree distribution as any study network, but unlike the method presented in Ref.~\cite{Miloetal2003}, the present process relies explicitly on network growth and can accommodate arbitrary node ``colorizations" (functional categories). Another standard control, the edge randomization of an arbitrary network, is easily implemented by setting $P_N=0$, $P_E=1$, $q=0.5$ (with $P_D=0$). This setting will remove and place edges randomly while keeping the total number of edges and nodes the same, resulting in a randomized graph after a sufficient number of updates. 

We have shown the a broad set of standard results in network analysis, concerning the edge distribution, modularity, global network structure, and critical behavior, can be reproduced in networks grown via a random process, with only a few tunable parameters. These networks, however, are entirely devoid of any function, and their properties are a consequence of the stochastic nature of the growth process only. We can conclude that while such properties may be useful for real-world networks, they are most likely not a consequence of the network's functionality, but rather a consequence of how they emerged.


\section*{Models}
\begin{figure}[!ht]
\begin{center}
\includegraphics[width=3.5in]{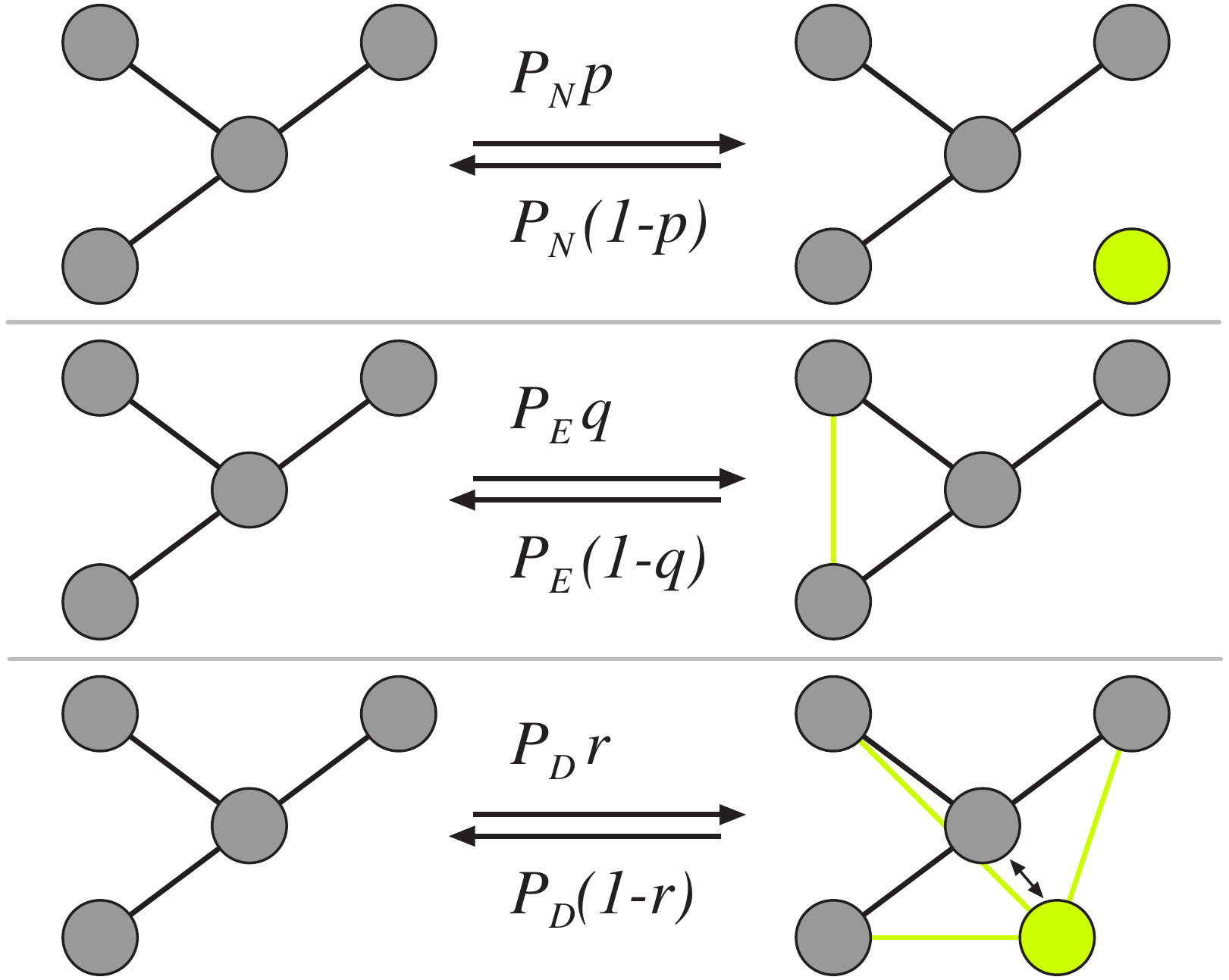}
\end{center}
\caption{
{\bf Network growth events and probabilities}.  Growth (or shrinkage) of network is determined by node, edge, and duplication events with specified probabilities, as described in the text.  
}
\label{model}
\end{figure}

Our networks are generated from one or several seed nodes. Subsequent {\em events}, determined by user-chosen probabilities, happen stochastically and usually lead to network growth. For example, the {\em node event probability} $P_N$ determines that a node is either added (without edges) or deleted, depending on a second parameter, the {\em node addition} probability $p$. Thus, for a single node event, the probability that a node will be added is $pP_N$ while the probability that a node will be deleted is $P_N(1-p)$ (see Figure~\ref{model}). A second type of event affects edges, with probability $P_E$: the {\em edge event probability}. Just as with nodes, edges will be added with an {\em edge addition probability} $q$, so that a single edge event will add an edge with probability $qP_E$ while an edge is removed with probability $(1-q)P_E$. Note that wile node addition or removal happens unconditionally, edge addition or removal is not guaranteed. The algorithm will only place an edge if the pair of nodes that is randomly selected is unconnected. Similarly, an edge removal instruction is only carried out if the pair of nodes that is randomly selected already has a connection, and otherwise fails. As a consequence, even edge addition probabilities $q<0.5$ will lead to a steady-state distribution of edges to nodes. 

We can calculate the steady-state distribution of edges per node (mean degree $\la k\ra$)  by calculating the total number of nodes $n$ and edges $m$ as a function of the number of events $N$ and the parameters $P_N$, $P_E$, $p$, and $q$ (we do not consider duplications in this calculation). The number of nodes added per event is $pP_N$ and the number of nodes removed is $(1-p)P_N$, so that the net number of nodes added after $N$ events
\be n=NP_N(2p-1)\;.
\ee
The net number of edges $m$ added is more complicated, because edges are only added with probability $q(1-\xi)$, and removed with probability $(1-q)\xi$ per event, where $\xi$ is the graph sparseness defined in Eq.~(\ref{xi}), and represents the probability that a random pair of edges is connected by an edge. At the same time, every time a node is removed, the algorithm removes the edges attached to it. On average then, a node removal even subtracts the average degree of that node, which is $\sum_i k_i/n=2m/n=\la k\ra $, so that
\be
m=NP_E(q-\xi)-NP_N\la k\ra (1-p)\;.
\ee
We can  then write an equation for the asymptotic dependence of the mean degree
\be
\la k\ra=\frac{2m}{n}=\frac{2P_E(q-\xi)-2P_N\la k\ra(1-p)}{P_N(2p-1)}
\ee
or
\be \la k\ra\approx\frac {q\eta}{1+\eta/ n}\;, \label{predic}
\ee
where $\eta=P_E/P_N$, and using $\xi\approx \la k\ra/n$, which holds for large $n$. Thus we see that in the limit of large $n$,
\be
\la k\ra\longrightarrow \eta q\;,
\ee
a behavior that is borne out in the simulations (data not shown).

A third type of event leads to node duplication or merger (fusion), controlled by the parameter $P_D$. Here, a node is duplicated with probability $r P_D$, while two nodes are fused with probability  $(1-r)P_D$ (the parameters are summarized in Table 1). While edge and node events are straightforward, node duplication/fusion events need more explanation because there are several different ways in which nodes can be duplicated or fused. Here, we implement an algorithm in which node duplication is directly related to the concept of modules: When duplicating a node, the new node is by definition in the {\em same module} as its ancestor, and the new node is connected to all nodes the ancestor is connected to. In order to implement this, nodes have to be assigned a {\em tag} that determines the module they belong to, the moment the node is created. To do this, the number of modules 
has to be given at the outset, and the probability for a node of color $k$ to connect to a node of color $\ell$ is obtained from the assortativity matrix $e$, which stores the fraction of edges between pairs of colors. This matrix was introduced earlier as Eq.~(\ref{ematrix}).  

For the case of node merging, two nodes (A and B) are picked at random. Node A keeps its connections and in addition obtains all the connections that node B had, upon which node B is deleted. The selection of the nodes to be merged is module independent, and thus could either merge nodes within a module or across modules. 

When growing modular networks, nodes are assigned a color based on a vector of probabilities that can be specified beforehand (for all the results in this paper, nodes are assigned to a module randomly at the time they are created). If an edge event specifies the placement of an edge, a random pair of nodes is selected and the identity of the colors determined. At this time, a random uniform number is drawn, and the edge is placed if this number is smaller than the corresponding probability in the $e$-matrix (\ref{ematrix}). If no edge is set, the algorithm tries to set the edge for a different pair of nodes, and attempts this up to 1,000 times. 

We determine a growth stop criterion either by specifying a maximum number of nodes, or by specifying a fixed number of iterations of the algorithm. In principle, the algorithm allows for the generation of both directed and undirected graphs. Here, we restrict ourselves to networks with undirected edges, and furthermore prevent nodes from connecting to themselves. Finally, two different growth initial conditions are possible: one in which we start with a single node from the first module in the $e$-matrix, the other where we start with a single node from every module. The $N_c(N_c-1)/2$ entries in the assortativity matrix, together with the number of colors and the six probabilities for stochastic network growth described above fully determine the structure of the network.

\section*{Supplementary Information}
\subsection*{Supplementary Text}
\subsubsection*{Structural and Functional 
Modularity}
Newman's modularity measure Eq.~(\ref{QN}) can be shown to be related to his assortativity measure (\ref{assort}) by noting that the mixing matrix $e$ is related to the adjacency matrix $A$ via a transform involving a matrix $F$ that relates nodes to modules:
\be
F_{ki}=\left\{\begin{array}{l}1\ \ \ \ {\rm if\ node}\  i\ {\rm in\ module\ }\ k\ \\
0\ \ \ {\rm otherwise}
\end{array}\right.\;.
\ee
Introduce
\be
E_{kl}=\sum_{ij}F_{ki}A_{ij}F^T_{jl}\;.
\ee
Because $\sum_k F_{ki}=\sum_l F^T_{jl}=1$, we find that
\be
||E||=\sum_{kl}E_{kl}=\sum_{ij}A_{ij}=2m\;,
\ee
where the notation $||...||$ indicates taking the sum of all the matrix elements. 
The mixing matrix $e$ is then just 
\be
e_{kl}=\frac{E_{kl}}{||E||}=\frac1{2m}\sum_{ij}F_{ki}A_{ij}F^T_{jl}\;.
\ee
Noting that ${\rm Tr}FF^T=S$, the modularity matrix defined above (\ref{QN}), we find that
\be
{\rm Tr}\ e=\frac1{2m}\sum_{ij}A_{ij}S_{ij}\;.
\ee

The same construction also allows us to write $Q_H$ [defined in Eq.~\ref{QH})] in terms of ${\rm Tr}\, e$ [Eq.~(\ref{QHandTre})] by noting that
\be
\tilde S_{ij}=(1+\frac1{N_c-1})S_{ij}-\frac1{N_c-1}{\mathbf 1}_{ij}
\;,
\ee
where ${\mathbf 1}$ is a matrix where each entry is 1. 

\begin{figure}[htbp] 
   \centering
   \includegraphics[width=2.5in]{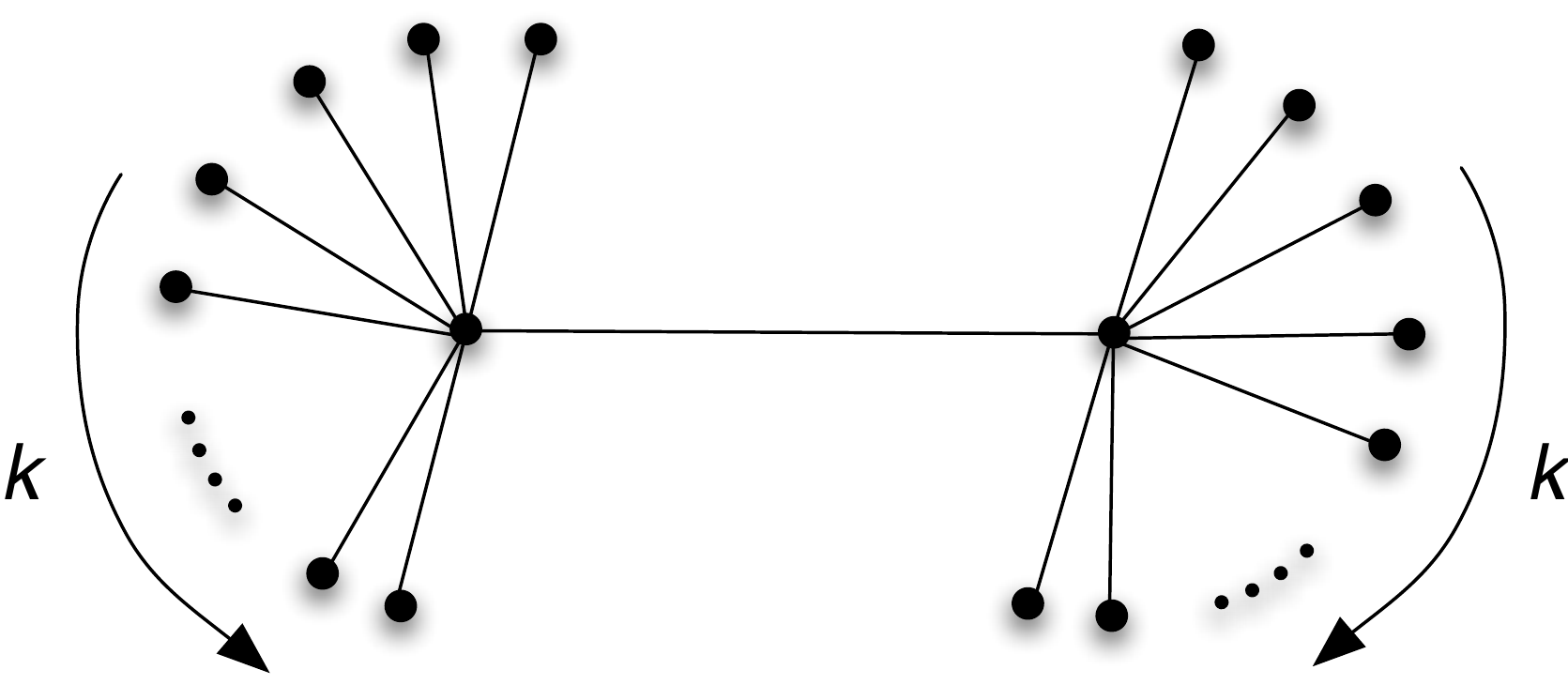} 
   \caption{Two-hub graph}
   \label{fig:extreme}
\end{figure}

We can test the limits of the modularity measures (\ref{QN}) and (\ref{QH}) by calculating the modularity of an extreme graph as depicted in Fig.~\ref{fig:extreme}, which is a graph of two hubs of degree $k$ connected by a single edge, and assuming that all of the nodes of one hub belong to the same module. In the limit of large $k$, such a graph should be classified as highly modular. However, Newman's measure applied to this graph gives
\be
Q_N=\frac12-\frac1{2k+1}\xrightarrow[k\rightarrow\infty]{}\frac12\;.
\ee
In comparison, the functional modularity measure $Q_H$, making the same assumptions about the modules, tends to 1 in the limit of high-degree hubs:
\be
Q_H=1-\frac32\frac1{2k+2}\xrightarrow[k\rightarrow\infty]{}1\;.
\ee

\subsubsection*{Functional modularity and assortativity}
Newman's assortativity $r$ [Eq.~(\ref{assort})] and the functional modularity measure $Q_H$ [Eq.~(\ref{QH})] are identical for a particular assortativity model, which we call the ``equal opportunity" model. It is defined by a probability $\pi$ of a color to connect to a node of the same color, but otherwise connects to other colors with equal probability:
\be
 e=\frac1{N_c}\left(  \begin{matrix} 
      \pi & \frac{1-\pi}{N_c-1}& \ldots& \frac{1-\pi}{N_c-1} \\
       \frac{1-\pi}{N_c-1} & \pi&\ldots& \frac{1-\pi}{N_c-1} \\
        \vdots&\vdots&\ddots& \vdots\\
         \frac{1-\pi}{N_c-1}& \frac{1-\pi}{N_c-1}& \ldots&\pi\\
   \end{matrix}\right)\;.
   \ee
The factor $1/N_c$, where $N_c$ is the number of colors, serves to normalize  $e$ such that $||e||=1$. It is easy to see that the fraction of nodes connected to type-$k$ nodes: $a_k=\sum_{\ell}e_{k\ell}=1/N_c$, so that (${\rm Tr}\,e=\pi$)
\be
r=\frac{{\rm Tr}\,e-\sum_ka_k^2}{1-\sum_k a_k^2}=\frac{\pi-1/N_c}{1-1/N_c}=\frac{N_c}{N_c-1}\left(\pi-\frac1{N_c}\right)\;,
\ee
which agrees with $Q_H$ defined in Eq.~(\ref{QH}) in the main text.

\section*{Acknowledgments}
We would like to thank Alpan Raval and Bj\o rn \O stman for discussions and comments on the manuscript. 
\subsubsection*{Author contributions} AH and CA conducted the research and wrote the manuscript.
\subsubsection*{Software availability} The program to grow the networks described in this article will be made freely available.
\subsubsection*{Funding} This work was supported by NSF's Frontiers in Integrative Biological Research program grant FIBR-0527023.
\subsubsection*{Conflicting interests}The authors have declared that no competing interests exist.

\bibliography{Growth}


\section*{Tables}

\begin{table}[!ht]
\caption{
\bf{Parameters of network growth model}}
\begin{tabular}{@{\vrule height 10.5pt depth4pt  width0pt} |c |p{150pt}|}
\hline
Abbreviation & probability \\ \hline
$P_N$& node event probability\\
$P_E$ & edge event probability \\
$P_D$ & Duplication event probability (duplication or fusion)\\
$p$ & Conditional node addition probability (given a node event). Conditional probability of node removal is $1-p$\\
$q$ & Conditional edge addition probability (given an edge event). $1-q$ is the conditional probability that an edge is removed\\
$r$ & Conditional node duplication probability (given a duplication/fusion event). Conditional node merging probability is $1-r$\\
\hline
\end{tabular}
\label{tab:label}
 \end{table}

\end{document}